\documentclass[12pt]{article}
\pdfoutput=1

\usepackage{cite}
\usepackage{booktabs}
\usepackage[english]{babel}
\usepackage{amsmath,amssymb,amsbsy,amstext, amsthm, simplewick}
\usepackage{hyperref}
\usepackage{graphicx}
\usepackage{amsfonts}
\usepackage{amssymb}
\usepackage[small]{caption}
\usepackage{upgreek}
\usepackage[svgnames,dvipsnames,x11names,table]{xcolor}
\usepackage{multirow}
\usepackage{geometry}
\usepackage[hang,flushmargin]{footmisc}
\usepackage{bm}
\usepackage{braket}
\usepackage{subcaption}
\usepackage{mathtools}
\usepackage{setspace}
\usepackage{cleveref}
\usepackage{comment}
\usepackage{scalerel}
\usepackage[normalem]{ulem}
\usepackage{slashed}
\usepackage{enumitem}
\usepackage{dsfont}
\usepackage{tikz}
\usetikzlibrary{decorations.markings}
\usetikzlibrary{shapes.misc}

\makeatletter
\g@addto@macro\bfseries{\boldmath}
\makeatother

\hypersetup{
    colorlinks=true,
    linkcolor={red!50!black},
    citecolor={blue!50!black},
    urlcolor={blue!80!black}
}

\usepackage{colortbl}

\makeatletter
\newlength{\apb@width}
\newcommand{\autoparbox}[2][c]{\settowidth{\apb@width}{#2}\parbox[#1]{\apb@width}{#2}}

\makeatother

\definecolor{lightgray}{gray}{0.9}

\usepackage[framemethod=default]{mdframed}
\newmdenv[skipabove=7pt,
skipbelow=7pt,
rightline=false,
leftline=false,
topline=false,
bottomline=false,
backgroundcolor=gray!10,
linecolor=gray,
innerleftmargin=5pt,
innerrightmargin=5pt,
innertopmargin=5pt,
innerbottommargin=5pt,
leftmargin=0cm,
rightmargin=0cm,
linewidth=4pt]{eBox}

\usepackage[most]{tcolorbox}
\tcbset{colback=white, colframe=black,
        highlight math style= {enhanced, 
            colframe=red,colback=red!10!white,boxsep=0pt}
        }
\definecolor{light-gray}{gray}{0.95}

\crefname{table}{Table}{Tables}
\crefname{equation}{Eq.}{Eqs.}
\crefname{appendix}{App.}{Apps.}
\crefname{section}{Sec.}{Secs.}
\crefname{figure}{Fig.}{Figs.}

\numberwithin{equation}{section}

\def\beq{\begin{equation}}
\def\eeq{\end{equation}}

\def\bea{\begin{eqnarray}}
\def\eea{\end{eqnarray}}

\def\zp{\zeta_{+}}
\def\zm{\zeta_{-}}
\def\zpm{\zeta_{\pm}}

\def\dzp{\dot{\zeta}_{+}}
\def\dzm{\dot{\zeta}_{-}}

\def\D{\nabla}

\def\beq{\begin{equation}}
\def\eeq{\end{equation}}
\def\bea{\begin{eqnarray}}
\def\eea{\end{eqnarray}}

\def\D{{\cal D}}

\def\O{{\cal O}}

\def\Mpl{M_{\rm pl}}

\def\k{{\vec{\scaleto{k}{7pt}}}}

\def\kp{{\!\!\vec{{\scaleto{\,\, k}{7pt}}^{\s\prime}}}}

\def\q{{\vec q}}
\def\p{{\vec p}}

\def\x{{\vec x}}

\def\y{{\vec y}}

\def\t{\texttt{t}}

\DeclareRobustCommand{\SkipTocEntry}[4]{}

\newcommand{\s}{\hspace{0.8pt}}

\definecolor{colorTC}{rgb}{.2,.7,.2}

\definecolor{amethyst}{rgb}{0.6, 0.4, 0.8}

\definecolor{acolor}{rgb}{0.4, 0.2, 0.4}

\definecolor{blue3}{RGB}{31, 119, 180}
\definecolor{red3}{RGB}{	214, 39, 40}
\definecolor{orange3}{RGB}{255, 127, 14}
\definecolor{green3}{RGB}{44, 160, 44}

\begin{document}

\begin{titlepage}
\setcounter{page}{1} \baselineskip=15.5pt
\thispagestyle{empty}
$\quad$
\vskip 70 pt

\begin{center}
{\fontsize{20.74}{24} \bf Soft Metric Fluctuations During Inflation}
\end{center}

\vskip 20pt
\begin{center}
\noindent
{\fontsize{12}{18}\selectfont Daniel Green and Kshitij Gupta}
\end{center}

\begin{center}
\vskip 4pt
\textit{{\small Department of Physics, University of California at San Diego,  La Jolla, CA 92093, USA}}

\end{center}

\vspace{0.4cm}
 \begin{center}{\bf Abstract}
 \end{center}

\noindent
The conservation of the long wavelength fluctuations of the metric plays a vital role in cosmology as the link between quantum fluctuations during inflation and late time observations. This is a well-known property of the classical evolution equations, but demonstrating that it is robust to quantum correction involves a number of technical arguments. In this paper, we will use effective field theory (EFT) techniques to demonstrate the all orders conservation of the super-horizon scalar and tensor fluctuations of the metric during inflation. We show how to construct an EFT for these soft modes, in analogy with Soft de Sitter Effective Theory. We pay particular attention to how the breaking of time-diffeomorphisms by the inflationary background introduces new time scales that alter the structure of the EFT. In this description, the all orders conservation of the metric fluctuations is a direct consequence of symmetries and power counting that cannot be altered by loop corrections. We further show that this holds when the inflaton (or metric fluctuations) is coupled to additional heavy fields, as in quasi-single field inflation. We match this behavior to several calculations in the ultraviolet (UV) theory and show how the Mellin representation enables a more transparent connection between the UV and the EFT descriptions.

\end{titlepage}
\setcounter{page}{2}

\restoregeometry

\begin{spacing}{1.2}
\newpage
\setcounter{tocdepth}{2}
\tableofcontents
\end{spacing}

\setstretch{1.1}
\newpage

\section{Introduction}

The seeds of structure in our Universe are well described by long-wavelength metric fluctuations produced during inflation~\cite{Baumann:2009ds,Achucarro:2022qrl}. The anisotropies in the cosmic microwave background and inhomogeneities in the distribution of matter encode the statistics of a scalar metric (adiabatic) fluctuation produced at much earlier times. Current observations are consistent with the predictions for these scalar fluctuations from many inflationary models~\cite{Green:2022bre}. Moreover, a coming generation of surveys will increase the sensitivity to CMB B-modes, which encode the signal of gravitational
waves (tensor modes) produced during the inflationary epoch~\cite{SimonsObservatory:2018koc,CMB-S4:2020lpa,CMB-S4:2022ght,Chang:2022lrw}.

The precise relationship between quantum fluctuations during inflation and the statistics of density fluctuations in the later universe is potentially sensitive to quantum corrections~\cite{Weinberg:2005vy,Weinberg:2006ac}. It is a remarkable feature of general relativity in accelerating cosmologies that the metric fluctuations do not evolve on long distances~\cite{Salopek:1990jq}. Importantly, this ensures a trivial relationship (classically) between the early and late universe in single field inflation. The conservation of the adiabatic mode is also deeply tied to the single field consistency conditions~\cite{Maldacena:2002vr,Creminelli:2004yq}, which offers one of the most powerful windows into the dynamics of inflation~\cite{Alvarez:2014vva}.

Despite its importance, the conservation of the metric fluctuations quantum mechanically rests on surprisingly technical arguments. It is generally appreciated that conservation is tied to the nonlinearly realized symmetries of the long wavelength modes~\cite{Weinberg:2003sw,Hinterbichler:2012nm,Hinterbichler:2013dpa}. Unfortunately, symmetry alone is insufficient to prove conservation and has to be supplemented with a variety of diagrammatic arguments~\cite{Pimentel:2012tw,Assassi:2012et,Senatore:2012ya}. Moreover, due to their complexity, these arguments have not been extended to the tensor modes (although similar arguments are expected to apply).

The central challenge with many approaches to loops in cosmology is that the expected size of a given correction is not easily determined. For example, although the action contains positive powers of the scale factor, $a(t)$, it was shown in~\cite{Weinberg:2006ac} that quantum corrections are, at most, logarithmic in $a(t)$. In this case, a detailed argument was used to show that the positive powers are always cancelled by inverse powers of the scale factor from the form of the propagators. Although much progress has been made in understanding the nature of loop corrections in de Sitter space and inflation~\cite{Antoniadis:1985pj,Tsamis:1994ca,Tsamis:1996qm,Adshead:2008gk,Senatore:2009cf,Burgess:2010dd,Senatore:2012nq,Akhmedov:2013vka,Anninos:2014lwa,Akhmedov:2019cfd,Gorbenko:2019rza,Mirbabayi:2019qtx,Baumgart:2019clc,Baumgart:2020oby,Benincasa:2022gtd,Benincasa:2024ptf}, the physical simplicity of the final results is not manifest in the mathematical framework for calculating cosmological correlators.

Effective field theory (EFT)~\cite{Manohar:2018aog,Cohen:2019wxr} is promising strategy for understanding the conservation of the metric fluctuation~\cite{Green2023}. The goal of defining an EFT is to make the size of corrections explicit (typically in the action or equations of motion) by choosing the appropriate variables and a renormalization scheme for the physical regime of interest. This approach has been successfully applied to long wavelength fluctuations of scalar fields in de Sitter space in Soft de Sitter Effective Theory (SdSET)~\cite{Cohen:2020php}. In this approach, one defines the degrees of freedom in the regime $k/(aH) \ll 1$ where $k$ is the length of the comoving wave-vector and $H$ is the Hubble parameter, so that the fields behave as scaling operators under rescaling of the coordinate distance (dilatations). In this framework, loop correlations take a familiar scaleless form and can be regulated using generalizations of dimensional regularization. The size of perturbative corrections organizes into powers of $k/(aH)$ that are manifest in the action. The interactions of light fields do generate large logs which can be resummed by conventional renormalization group techniques to give the framework of Stochastic Inflation~\cite{Starobinsky:1986fx,Nambu:1987ef,Starobinsky:1994bd} (at leading order) and corrections thereof~\cite{Cohen:2021fzf,Cohen:2021jbo,Cohen:2022clv}. 

Our goal in this paper is to write the EFT for soft metric fluctuations during inflation. This presents two non-trivial challenges beyond SdSET. First, the inflationary background breaks the de Sitter isometries, allowing for explicit time dependence in the action and for non-Lorentz invariant interactions~\cite{Cheung:2007st}. Second, the action of the canonical metric fluctuations, after gauge fixing and imposing constraints, is not manifestly local~\cite{Maldacena:2002vr}. While it is desirable to write down the EFT for these degrees of freedom directly, the possibility of non-local terms adds some nuance to the usual EFT playbook. 

The appeal of this approach is that it will make the conservation of both the scalar and tensor modes of the metric manifest from power counting and symmetries. Ultimately, symmetries allow us to circumvent the most acute challenges of working with the metric fluctuations during inflation. They define a robust set of observables~\cite{Green:2020ebl} that are subject to non-trivial Ward identities~\cite{Creminelli:2012ed,Assassi:2012zq,Goldberger:2013rsa,Pimentel:2013gza}, without appealing to locality or de Sitter invariance. These symmetries dictate that one of the EFT degrees of freedom is time-independent in the free EFT, and that interactions decouple at late times (are formally irrelevant). This holds, not only in single field inflation, but also for couplings of the inflaton or metric to additional fields that redshift away, such as quasi-single field inflation~\cite{Chen:2009zp,Baumann:2011nk,Noumi:2012vr}. As a result, one can see directly from the construction of the EFT that conservation of the metric fluctuations survives quantum corrections, much like how Wilsonsian RG and EFT clarify the meaning of renormalizability in terms of dimensional analysis~\cite{Polchinski:1983gv}.

Like SdSET, matching to the UV description requires determining both the initial conditions and the couplings of the effective action. We use the Mellin representation~\cite{Sleight:2019mgd,Sleight:2019hfp,Premkumar:2021mlz,Qin:2022lva,Qin:2023bjk,Cohen:2024anu} of the full one-loop power spectrum to demonstrate how the EFT matching connects to the full microscopic calculation. Mellin space has the useful property of organizing the UV calculation into an infinite sum over power-laws, which makes the connection to the EFT transparent. This approach also allows us to directly compare (and contrast) the single field inflation results with the non-trivial anomalous dimensions associated with heavy fields. Principal series fields are known to have anomalous dimensions~\cite{Bros:2010rku,Marolf:2010zp,Hogervorst:2021uvp,Chakraborty:2023qbp,Chakraborty:2023eoq}, and arise from non-dynamical operators that must be added to the EFT~\cite{Green:2023ids,Cohen:2024anu}. We show that no such operators are needed in single field inflation, thus ensuring conservation follows from power counting.

This paper is organized as follows: In Section~\ref{sec:EFT}, we will introduce the EFT of the soft scalar metric fluctuations and how it is organized by symmetries and renormalization. In Section~\ref{sec:all_orders}, we will use power counting of the EFT to demonstrate all orders conservation. We show explicitly how this arises at one-loop in the UV theory by matching. In Section~\ref{sec:tensors}, we extend the EFT and all orders conservation results to include tensor modes. We conclude in Section~\ref{sec:conclusions}. This paper includes three appendices: In Appendix~\ref{app:EFT}, we include details of the derivation of the EFT. Appendix~\ref{app:mellin} details the one-loop matching calculation. Finally, Appendix~\ref{app:flat} provides a flat space example of the scheme dependence of one-loop calculations.

\section{EFT of Scalar Modes}\label{sec:EFT}

We will start by working out the basic structure of the EFT of the soft scalar mode of the metric during single-field inflation. Like SdSET, much of the structure can be understood starting from the classical description of the metric in the limit of vanishing wavenumber. We will then use this classical intuition to choose an ansatz for the field operators in the soft limit and construct the action of the EFT.

\subsection{Inflationary Background}

When one thinks of EFT and single-field inflation, it is natural to start from the EFT of Inflation~\cite{Creminelli:2006xe,Cheung:2007st} as our UV starting point. In this description, the homogeneous universe is described by an FRW metric $ds^2 =-dt^2 + a^2(t) d\x^2$, where the Hubble parameter $H(t)$ obeys
\beq
H(t) =  \frac{\dot a(t)}{a(t)} \qquad  - \dot H(t) \ll H^2(t) \ .
\eeq
Here $H(t)$ is some function determined by the unknown UV theory. 

During inflation, the background field defines a preferred time slicing where each spacelike surface is defined by constant field, $\phi(\x,t) = \phi(t)$. Furthermore, the evolution of the field is approximately linear in time, $\phi \propto t$ so that time translations are broken by the background. In the EFT of Inflation, one describes the fluctuations around this background by introducing the Goldstone boson $\pi$ associated with the broken time translations (in the flat space limit). It is  useful to define the action in terms of  
\beq
U(\x,t) = t +\pi(\x,t) \ ,
\eeq 
where $U(\x,t)$ transforms as a scalar under diffeomorphisms, but $\pi(\x,t)$ transforms nonlinearly. This field is coupled to a dynamical metric, now in the ADM form
\beq
ds^2 = - N dt^2 + a(t)^2 [e^{2 \zeta + 2 \gamma}]_{ij} (dx^i + N^i dt) (dx^j + N^j dt)  \ .
\eeq
The most general possible action for $\pi$ can be constructed treating $U$ as a scalar field and then expanding in $\pi$ and the metric components~\cite{Baumann:2011su}. For calculating correlators of the scalar metric fluctuation, $\zeta$, at horizon crossing, it is advantageous to use the gauge where $\zeta =0$ and the Goldstone boson equivalence theorem~\cite{Green:2022slj,Green:2024hbw} ensures we can neglect dynamical gravity. One can then return to the variable $\zeta$, from $\pi$ via the transformation
\beq
\zeta = - H \pi +\ldots \ .
\eeq
This is sufficient to calculate the leading observational signal of a wide range of inflationary models.

However, the long distance behavior of an inflating universe is not well described in terms of $\pi$, but instead is more manifest in terms of $\zeta$ and $\gamma$ directly~\cite{Salopek:1990re,Wands:2000dp}. For example, the fact that $\zeta(\k, t)$ and $\gamma_{ij}(\k,t)$ are constant when $k \ll aH$ is an essential ingredient used across cosmology when relating the observed maps of the universe to the dynamics during inflation~\cite{Salopek:1990jq}. This observation can be traced to the statement that $\zeta$ non-linearly realizes the group of dS isometries~\cite{Creminelli:2012ed,Hinterbichler:2012nm}
\beq
\begin{aligned}
\x &\to e^\lambda \x \qquad \zeta(\x,t) \to \zeta(e^\lambda \x,t) + \lambda \\[0.1cm]
\x &\to \y = \x + \vec {b} (e^{-2Ht} H^{-2} - \x^2 ) + 2 (\vec{b} \cdot \x ) \x  \qquad \zeta(\x, t) \to \zeta(\y, t) - 2 \vec{b} \cdot x\ .
\end{aligned}
\eeq
These symmetries imply the inflationary correlators obey a set of Ward identities, known as the single-field consistency conditions. These conditions can be directly implemented at the level of late time observables.

Our goal in this section is to write down an EFT for $\zeta$ directly, holding $\gamma_{ij} = 0$ fixed for now. As a starting point, we will use the action for $\zeta$ derived from the top down~\cite{Chen:2006nt,Cheung:2007sv}, before understanding the same behavior from the bottom up.

\subsection{EFT from the Top Down}

We will start with the quadratic action for the full adiabatic mode $\zeta$ 
\beq
S_2 = \int dt d^3 x  a^3(t) \frac{\Mpl^2 \dot H(t)}{c_s^2(t) H^2(t)}\left(- \dot \zeta^2 +\frac{ c_s^2(t)}{a^2(t)} \partial_i \zeta \partial^i \zeta \right) \ ,
\eeq
where we recall we can assume $\dot H < 0$ by enforcing the null-energy condition. Our goal here is to write an EFT that isolated the physics of the long wavelength modes, namely those with $c_s(t) k \ll aH(t)$. To understand the behavior in this limit, we can solve the equations of motion taking $k = 0$
\beq
\frac{\partial}{\partial t} \left( a^3 \frac{\dot H(t)}{c_s^2(t)H^2(t)} \dot \zeta \right)  = 0 \to \zeta =c_0 + c_1  \int^t dt' \frac{c_s^2(t')H^2(t')}{a^3 \dot H(t') }
\eeq
To get intuition, it will occasionally be more useful to work around a specific time $t=t_\star$ (or conformal time $\tau_\star$) and treat the time evolution as power law
\beq
H(t) = H(t_\star)  \left( \frac{a(t)}{a(t_\star)} \right)^{-\epsilon(t_\star)} \qquad \dot H(t) = - \epsilon(t_\star) \left(\frac{a(t)}{a(t_\star)} \right)^\eta H^2(t)
\eeq
where we defined the slow roll parameters
\beq
\begin{aligned}
\epsilon = -\frac{\dot H}{H^2} \qquad \eta =\frac{\dot{\epsilon}}{\epsilon H} \qquad
s =\frac{\dot{c_s}}{c_s H}
\end{aligned}
\eeq
We can then treat $\zeta$ as an approximate power law for $t \sim t_\star$
\beq
\zeta(\k= 0,t) = c_0 + c_1 \left(\frac{a(t) H(t)}{c_s(t)}\right)^{-\beta} \qquad \beta = 3+ 2 \epsilon + \eta + s + {\cal O}(\epsilon^2) \ .
\eeq
where ${\cal O}(\epsilon^2)$ includes all terms quadratic in the slow roll parameters $\eta, s$ and $\epsilon$. When necessary, these logs that arise from the slow-roll expansion can be resummed~\cite{Baumann:2014cja}. Here we have written our solution in terms of $a(t) H(t)/c_s(t)$ in anticipation of the EFT expansion in $c_s(t) k / [a(t) H(t)] \ll 1$.

The slow-roll expansion is a natural starting point to derive an EFT for the soft modes, in direct analogy with SdSET. Since the time dependence is nearly the same as dS, we factor our the power-law behavior in time 
\beq
\zeta(\x,t) = \zp(\x,t) + \left(\frac{a(t) H(t)}{c_s(t)} \right)^{-\beta} \zm(\x,t)  \equiv  \zp(\x,t) + \Lambda(t)^{-\beta} \zm(\x,t)\ ,
\eeq
where we defined the (comoving) UV cutoff as $\Lambda(t) \equiv a(t) H(t)/ c_s(t)$. At zeroth order in slow-roll, $\beta = 3$ and $H(t) = H(t_\star)$ and this is the same ansatz as a massless scalar in dS. Using
\beq
\dot \zeta = \dzp -\beta H(t)(1-\epsilon-s) (a(t) H(t))^{-\beta} \zm + (a(t) H(t))^{-\beta}\dzm \ ,
\eeq
we can substitute into the quadratic action to find
\beq
S = \int dt d^3 x a(t)^3\frac{\Mpl^2 | \dot H(t) |}{c_s^2(t)H^2(t)} \left[ \dzp^2  + 2\beta (1-\epsilon-s) H (aH)^{-\beta} \dzp \zm + (aH)^{-2\beta} \dzm^2 +\ldots\right] 
\eeq
where we have dropped the terms suppressed by $k^2/a^2$ or higher. Notice that, by definition,
\beq
\frac{\partial}{\partial t} \left( a^3\frac{ \dot H(t)}{c_s^2(t)H^2(t)}\beta (1-\epsilon-s) H (aH)^{-\beta} \right)  = 0 \ ,
\eeq
So that the coefficient of the dominant kinetic term is time independent. In fact, this is a completely general consequence of our ansatz and not only a statement at low orders in the slow-roll expansion.

To see how the EFT will behave even away from the slow-roll limit, we define coefficient of the kinetic term $\kappa(t)$ and the non-trivial solution $\rho(t)$ as
\beq
\kappa(t) \equiv a(t)^3\frac{\Mpl^2 | \dot H(t)|} {c_s^2(t) H^2(t)}  \qquad 
\frac{\partial}{\partial t} \left( \kappa(t)  \dot \rho(t) \right)  = 0 \ .
\eeq
We will take the ansatz
\beq
\zeta(\x,t) = \zp(\x,t) + \rho(t) \zm(\x,t)
\eeq
which also implies
\begin{equation}
	\dot{\zeta} = \dzp + \dot{\rho} \zm + \rho \dzm \qquad \partial_i \zeta = \partial_i \zp + \rho \partial_i \zm \ .
\end{equation}
From the top-down, we may simply substitute this ansatz into the quadratic action,
\begin{equation}
	S = \int dt \, d^3x\, \kappa(t) \bigg[ \dot{\zeta}^2 - \frac{c_s^2}{a^2} (\partial_i \zeta)^2 \bigg] \ ,
\end{equation}
to determine the quadratic action of the EFT. This is given by
\begin{equation}
\begin{aligned}
	S &= \int dt\, d^3x\, \kappa(t) \bigg[ (\dzp + \dot{\rho} \zm + \rho \dzm)^2 - \frac{c_s^2}{a^2} (\partial_i \zp + \rho \partial_i \zm)^2  \bigg]\\
	&= \int dt d^3x\, \kappa(t) \bigg[ 2 \dot{\rho} \dzp \zm + 2 \rho \dzp \dzm +\dzp^2 + \rho^2 \dzm^2- 2 \rho \frac{c_s^2}{a^2} \partial_i \zp \partial_i \zm \bigg]
\end{aligned}
\end{equation}
Where we eliminated the other terms by the same integration by parts and field redefinitions as in SdSET, but now with the equation of motion $\dot \kappa \dot \rho + \kappa \ddot \rho =0$. We then find that leading kinetic term is $2\kappa\dot \rho \dzp \zm$, where $\kappa(t)\dot \rho$ is a constant (time-independent) by construction. Therefore, the leading order equations of motion become
\begin{equation}
	\partial_t (\kappa(t) \dot{\rho} \zp )= 0
\end{equation}
However, we also have second order kinetic terms. We can remove $\dzp^2$ by a field redefinition of $\zm$ and we can neglect $\dzm^2$ because it is suppressed by $\rho^2$ and can be ignored.  However when computing the ${\cal O}(k^2)$ correction, we cannot drop the  $\dzp \dzm$ term. Instead, we must treat is a perturbation so as to not introduce additional (unphysical) degrees of freedom. The resulting interaction Hamiltonian becomes, so that
\begin{equation}
    H_{\rm int} =  \int d^3x\, \bigg[ - 2\rho(t) \kappa(t) \dzp \dzm + 2 \kappa \rho \frac{c_s^2}{a^2} \partial_i \zp \partial_i \zm \bigg]
\end{equation} 
The result evolution of $\zp$ at subleading order is given by 
\beq
\begin{aligned}
    \zp^{(2)} &= i \int dt \, d^3x\, \bigg[- 2\rho(t) \kappa(t) \dzp [\dzm, \zp] + 2 \kappa \rho \frac{c_s^2}{a^2} \partial_i \zp [\partial_i \zm, \zp] \bigg]\\
    &= i \int dt \, \frac{1}{2 \kappa \dot{\rho}} \bigg[ -2 \partial_t (\rho \kappa \dzp ) + 2 \kappa \rho \frac{c_s^2}{a^2} \partial^2 \zp  \bigg]
\end{aligned}
\eeq
where $\zpm$ on the RHS is just the interaction picture operator. Specializing to the case of slow-roll inflation, at leading order in the slow roll parameters, this expression gives
\bea
    \dzp = \frac{i}{3} \left( -\frac{\ddot{\zeta}_+}{H} + \frac{c_s^2}{a^2 H} \partial^2 \zp \right)
\eea
We emphasize here that $\dzp \dzm$ gives a correction at order $k^2$, that is needed to get the correct result. We can understand this result as the statement that we should use the $\zp$ equations of motion to eliminate higher time derivatives, but that we need to be more careful of higher time derivatives of $\zm$. These contributions are always subleading and can usually be addressed by integrating by parts first.

The canonical momentum defined with respect to $\zp$ is given by the action to be 
\beq
\Pi = 2\kappa \dot \rho \zm + {\cal O}\left(\frac{k^2}{a^2}\zp \right)
\eeq
and therefore a $k \to 0$, we have
\beq
[\zp(\x,t),\zm(\x,t)] = \frac{i}{2 \kappa \dot \rho} \delta(\x-\x') \to[\zp(\k,t),\zm(\kp,t)] = \frac{i}{2 \kappa \dot \rho} (2\pi)^3 \delta(\k+\kp) 
\eeq
As with SdSET, the commutator being ${\cal O}(k^0)$ fixed the relation between the scaling with $k$ of $\zp$ and $\zm$.

The initial conditions for $\zp$ when it enters our EFT, at horizon crossing, are not determined by the EFT itself but are fixed by matching. In the free theory, since $\zeta \to \zp$ as $t\to \infty$, we can match the power spectrum to find
\beq
\langle \zp(\k) \zp(\kp) \rangle' = \frac{H(t_\star)^4}{4 \Mpl^2 \dot H(t_\star) c_s(t_\star)} \frac{1}{k^3} \left(\frac{k_\star}{k}\right)^{2 \epsilon + \eta + s} \equiv  \frac{\Delta_\zeta^2}{k^3} \left(\frac{k}{k_\star}\right)^{n_s-1}
\eeq
The tilt of the spectrum, $n_s -1 = -2\epsilon - \eta -s$, it directly related to $3-\beta  =n_s-1$. This is not a coincidence, as can be understood as follows: in the $k / \Lambda(t)\to 0$ limit, we might anticipate the mode functions take the form 
\beq\label{eq:modefnc_guess}
\bar \zeta \approx \frac{\Delta_\zeta}{k^{(3 -(n_s-1))/2}}\left(\left(1 + {\cal O}(k^2/\Lambda^2)\right) + i \left(\frac{k}{\Lambda(t)} \right)^\beta\left(C+{\cal O}(k^2/\Lambda^2)\right)  \right) \ .
\eeq
This ansatz is based on the assumption that the corrections to the leading scaling behavior can be expressed as an expansion in $k/\Lambda(t)$. With this assumption, if we write the $\zeta$ operator as $\hat \zeta = \bar \zeta \hat a^\dagger+ {\rm h.c.}$, then the commutator must be
\beq
[\hat \zeta, \hat {\dot \zeta}] \propto i k^{-3+(n_s-1) +\beta} \ .
\eeq
However, since the commutator must vanish at spacelike separation, we know that this  must be a contact term (scale as $k^0$) and therefore
\beq
-3 + (n_s-1) + \beta = 0 \to (n_s-1) = 3-\beta \ .
\eeq
In this precise sense, the organization of corrections into powers of $k/\Lambda(t)$ is consistent with the scaling of $\zp$ and $\zm$, and is matched by the result of an explicit calculation. Furthermore, the ansatz in Equation~(\ref{eq:modefnc_guess}) can be justified more rigorously from approximate symmetries of the UV description~\cite{Green:toappear}.

\subsection{Bottom-Up EFT}

The goal of writing an EFT for the soft modes during inflation is to make transparent interactions in the theory alter the long wavelength behavior. As with any EFT, we want to write down the most general action consistent with the symmetries. From there, we expect that the relative important of different terms is fixed by power counting and renormalization. The power counting parameter is the (comoving) soft momentum $k$ compared to scale of horizon crossing
\beq
\Lambda(t) =\frac{a(t) H(t)}{c_s(t)} \quad \to \quad \frac{k}{\Lambda(t)} \ll 1 \ .
\eeq
Like any EFT, we will use the UV scale $\Lambda(t)$ to make the action dimensionless and, as a result, make power counting of the terms in the action explicit. 

The most important symmetry to single field inflation, is the dilation symmetry
\beq
\x \to e^\lambda \x \qquad \zeta(\x,t) \to \zeta(e^\lambda x,t) + \lambda 
\eeq
If we split $\zeta$ in to the components $\zpm$, then we see that $\zp$ transforms non-trivially
\beq
\zp(\x,t) \to \zp(e^\lambda x,t) + \lambda \qquad \zm(\x,t) \to \zm(e^\lambda x,t)
\eeq
Imposing this symmetry, $\zp$ must appear with derivatives or in power of $e^\zp$ that are fixed by scaling. For example, our leading acting might contain terms of the from
\begin{align}
S = \int dt d^3 x e^{3 \zp} &\bigg( \kappa\dot \rho \dzp \zm - \kappa(t)\rho(t) \Lambda(t)^{-2} e^{-2 \zp}\partial_i \zp \partial^i \zm \nonumber\\
&+\sum_n c_n(t) \rho(t) (\Lambda(t)^{-2} e^{-2\zp} \partial^2 \zp)^n \zm + \ldots   \bigg)
\end{align}
We have not included time-derivatives of $\zp$, given that the leading equations of motion allow us to relate $\dzp = a^{-2} \partial^2 \zp +\ldots$. This substitution clearly holds in perturbation theory, but can be made precise by a field redefinition.

Without any additional information, there is no particular response to keep these terms and not the many other possible terms. For the application to a wide variety of inflationary models, we can work within the slow-roll expansion, such that $\rho \approx (a(t) H(t))^{-3}$ and the time dependence of the couplings is suppressed, e.g.~$\dot c_n(t) / c_n(t) \ll 1$. 

When the slow roll parameters are small, we can power count the interactions treating $\zm$ as an operator with dimension $\beta \approx 3$. In this regime, power counting is effectively identical to SdSET with the added constraint of the non-linearly realized symmetries. To understand the scaling, we recall that if we rescale $x\to \lambda^{-1} x$, $\k \to \lambda \k$ and $\Lambda(t) \to \lambda \Lambda(t)$, the physical wavelengths are all held fixed in units of the Horizon radius. Therefore, if we count powers of $k$, we can restore units with powers of $\Lambda(t)$ up to the time-dependence of the couplings. Defining the dimension of $\zp$ and $\zm$ to be $\Delta_+ = (n_s-1)/2$ and $\Delta_- = (n_s-1)/2+\beta= 3 - (n_s-1)/2$, then a given interaction term can be written
\beq
S_{\rm int} = \int d^3 x dt c_n(t) (\Lambda)^{3 -n(\Delta_+ + 1)- m \Delta_-} \D(e^\zp \partial \zp)^n \zm^m
\eeq
Here the power of $3$ comes from power counting the measure $d^3 x$ as $k^{-3}$ and $dt$ as $k^0$. Like SdSET, we can remove any term with $m =0$ with a field redefinition of $\zm$. This can also be understood in perturbation theory since $[\zp,\zp ] =0$, we require at least one power of $\zm$ to generate a correction to a $\zp$ cosmological correlator.  As a result, the lowest dimension (local) operators will be
\beq
S_{\rm int} = \int d^3 x dt c_n(t) (\Lambda)^{-n-(n-1)(n_s-1)/2} (e^{-\zp} \partial \zp)^n \zm \ .
\eeq
The contribution of this interaction is proportional to $c_n(t) (k/\Lambda(t))^{-n -(n-1)(n_s-1)/2}$ and is therefore suppressed. In the language of RG flow, these interactions are all irrelevant by power counting and decouple at late times.

\subsection{Non-Local Terms}

Unfortunately, the assumption that our action is local is not justified in the case of metric fluctuations. As a result, we must ensure that there are no non-local terms in the action that alter the power counting conclusions. Concretely, when writing the metric in the ADM form,
\beq
ds^2 = - N^2 dt^2 + a(t)^2\delta_{ij} e^{2 \zeta} (d x^i + N^i dt) (d x^j N^j dt) \ ,
\eeq
the action is only manifestly local in terms of $N$, $N^i$, and $\zeta$.  However, $N$ and $N^i$ and non-dynamical and, at linear order in $\zeta$, the obey the constraints~\cite{Chen:2006nt}
\beq
\delta N  =  - \epsilon \zeta  \qquad N^i   = -\frac{\partial^i \zeta}{H} +\frac{\epsilon}{c_s^2 } \frac{\partial^i}{\partial^2} \dot\zeta \ .
\eeq
As a result, from local terms in the action in terms of $N$, $N^i$ and $\zeta$, we will generate non-locals for the $\zeta$ action alone, 
\beq\label{eq:nonlocalEG}
{\cal L} \supset \partial_j N^j N^i \partial_i \zeta \to \frac{\epsilon^2}{c_s^4} \dot \zeta \left(\frac{\partial_i}{\partial^2} \dot \zeta \right) \partial^i \zeta \ .
\eeq
Using our ansatz, this would generate an interaction in the EFT
\beq
{\cal L}_{\rm int} \supset c_3(t) \zm \left(\frac{\partial_i}{\partial^2} \dzp \right) \partial^i \zp
\eeq
This term in non-local at face value. However, we should recognize that the quadratic equation for motion for $\zp$ are 
\beq
\dzp = c_s^2(t) \frac{k^2}{a^2(t)} \zp + {\cal O}\left(\frac{k^2}{a^2} \right) \ .
\eeq
From experience with other EFTs, this should suggests that we can perform a field redefinition that effectively replaces $\dzp$ with its solution to the equation or motion. Concretely, if we have
\bea
S \supset \int d^3 x dt \zm (\dzp + c_s^2(t) \frac{\partial^2}{a^2} \zp) + c_3(t) \zm \left(\frac{\partial_i}{\partial^2} \dzp \right) \partial^i \zp \ ,
\eea
then under the change of variables
\beq
\zm \to \zm + c_3(t) \frac{\partial_i}{\partial^2}\left( \zm \partial^i \zeta\right) \ ,
\eeq
we get
\beq
\begin{aligned}
S \to \int d^3 x dt &\bigg( \zm (\dzp + c_s^2(t) \frac{\partial^2}{a^2} \zp) \\
&+ c_3(t) \zm \left(c_s^2(t)\frac{\partial_i}{a^2(t)} \dzp \right) \partial^i \zp + \frac{\partial_i}{\partial^2} \left( c_3(t) \zm \dzp \partial^i \zp \right)  \bigg) \ .
\end{aligned}
\eeq
While it is not obvious that the last term can be treated as a total derivative, the sum over in the individual terms vanishes by momentum conservation (for non-zero internal momenta).

The more general question that we must answer in an EFT is what non-local terms are allowed by symmetries and physical principles. While it may be clear $\dot \zeta$ is allowed by the shift symmetry $\zeta \to \zeta +\lambda$, it is less clear that $\nabla^{-2}  \dot \zeta$ would be. To address this question, we can define the charge under the dilation current using the conjugate momentum
\beq
\Pi = \zm  \to Q = \int d^3 x \, \delta \zeta \Pi 
\eeq
For this symmetry to leave the action invariant, we require $[Q, S] = \delta S = 0$. We write the action in terms of individual invariant operators so that 
\beq
S\supset \int d^4 x \sqrt{g} \sum_n \O_n(x) \to [Q,\int d^4 x \sqrt{g} \sum_n \O_n(x)] = 0
\eeq
Any potentially marginal operator, $\O_m$, will take that form $\O_m = \zm F(\zp,\partial_t, \vec \partial)$ where $F(\zp,\partial)$ may be non-local in $\vec \partial$. Now we can Fourier transform the fields so that 
\beq
\int d^4x \sqrt{g} \O_m \to c_m(t) \zm(\p_0) \left( \prod_{i=1}^N \int d^3 p_i \zp(\p_i) \right) \tilde F(\{ p_i \} ) (2\pi)^3 \delta(\sum_{n=0}^n \p_{n} )
\eeq
Since the shift symmetry arises from $Q= \gamma \zm(\q=0) + {\cal O}(\zp\zm)$, a single commutator will set one of the momenta of $\zp(\p_i)$ to zero, 
\beq
[Q, \zp(\p)] \to \gamma i (2\pi)^3 \delta(\p) +{\cal O}(\zp) \ .
\eeq
Therefore, after $N$ commutators we have 
\beq
[Q,[Q,...[Q,S],...,],] = 0 \supset \zm(\p_0=0) F(\{\p_i\} = 0 ) + {\cal O}(\zm \zp)   \ ,
\eeq
where $\{p_i\} = 0$ means all the individual momenta vanish. In the perturbative regime, the first term cannot be cancelled by higher orders in $\zp$ and therefore must vanish. Furthermore, by defining the charges at finite volume and taking the limit $V \to \infty$, we require that 
\beq
\lim_{\{\p_i\}\to 0 } F(\{\p_i\}) =0  \ ,
\eeq
for any order of limits. This is equivalent to the requirement that there are more powers of $\q$ in the numerator than the denominator. This is sufficient to ensure that any such operator is irrelevant by power counting.

For a concrete example, let us take the non-local term in Equation~(\ref{eq:nonlocalEG}) that arises from the leading ADM constraints. We should be able to understand the scaling dimension of the operator without any matching, based on the combination of $\{\zp, \zm, \partial\}$ that are present, up to slow roll factors. Using the EOM $\dzp = a^{-2}\partial^2 \zp$, we get
\begin{equation}
	\dot \zeta \left(\frac{\partial_i}{\partial^2} \dot \zeta \right) \partial^i \zeta  \supset \partial^2 \zp \partial_i \partial^{-2} \zm \partial_i \zp
\end{equation}
Up to slow roll factors, the dimension of the operator should be $-2$, and will hence be irrelevant. Still, we want to see how this works from our symmetry based argument. Let's therefore define the operator in the action
\begin{equation}
\begin{aligned}
    {\cal O}(x) &= c_1 \partial^2 \zp(x) \partial_i \partial^{-2} \zm(x) \partial_i \zp(x)\\
	 \to {\cal O}(k) &= c_1 \int \frac{d^3q_1}{(2\pi)^3} \frac{d^3q_2}{(2\pi)^3} \frac{q_1^2 q_{2,i}(k - q_1 - q_2)_i}{(k - q_1 - q_2)^2} \zp(q_1) \zp(q_2) \zm( k - q_1 - q_2)
\end{aligned}
\end{equation}
Now, let us act with the charge operator on ${\cal O}(k)$ defined at $k = 0$ (any choice of momentum is allowed). We get 
\begin{equation}
     [Q, {\cal O}(k = 0)] = c_2 \lim_{q_2 \to 0} \frac{d^3q_1}{(2\pi)^3}  \frac{q_1^2 q_{2,i}(k - q_1 - q_2)_i}{(k - q_1 - q_2)^2} \zp(q_1)  \zm( k - q_1 - q_2)
\end{equation}
With another commutator with the charge operator, we get, 
\begin{equation}
	[Q,[Q, {\cal O}(k = 0)]] = c_3 \lim_{q_1, q_2 \to 0}   \frac{q_1^2 q_{2,i}(k - q_1 - q_2)_i}{(k - q_1 - q_2)^2} \zm( k - q_1 - q_2) 
\end{equation}
Now, on taking this limit, we see that we smoothly get 0, since we have more factors of momenta in the numerator than denominator. In general, this condition will be satisfied if and only if we have more factors of momenta in the numerator than in the denominator. Since the physical momenta is $q/a$, this means we must have more factors of $a$ in the denominator than in the numerator, up to slow roll corrections, and hence the non-local terms are irrelevant.

\subsection{Renormalization}

As with SdSET and other perturbative calculations in cosmology, it is dynamical renormalization plays a non-trivial role organizing the perturbative calculation and resumming large logarithmic corrections. In pure dS (SdSET), every factor of $k$ must appear in the combination $k/[aH]$ and this procedure is straightforward. Specifically, we regulate divergences by analytic continuation of the dimensions~\cite{Cohen:2021fzf} so that we encounter divergences of the form
\bea
\langle\mathcal{O} \ldots\rangle &\supset& \langle \mathcal{O} ... \rangle_0 \times [a H]^{-2 \alpha} \int \frac{\mathrm{d}^3 p}{(2 \pi)^3} \frac{1}{\left(p^2+K^2\right)^{3 / 2-\alpha}} \\
& \rightarrow&\langle \mathcal{O}' ... \rangle \left(-\frac{1}{2 \pi^2}\left(\frac{1}{2 \alpha}+\log \frac{K}{[a H]}-\log 2\right) \right) \ ,
\eea
where $K$ is an IR regulator typically given by the (soft) external momenta. We then define renormalized operator by introducing a reference time $t_\star$and scale $K_\star$ so that
\beq
\mathcal{O}=Z_{\mathcal{O}} \mathcal{O}_R \quad \text { with } \quad Z_{\mathcal{O}}-1 \propto-\frac{1}{2 \pi^2}\left(\frac{1}{2 \alpha}+\log \frac{K_{\star}}{[a H]_{\star}}\right)
\eeq
where $[aH]_\star= a(t_\star)H$. As a result, the correction to the renormalized operator is under control, so that
\beq
\left\langle\mathcal{O}_R \ldots\right\rangle \propto \left\langle\mathcal{O}_R \ldots\right\rangle_0 \times \left( \frac{1}{2 \pi^2} \log \frac{[a H]_{\star}}{[a H]}+\log \frac{K}{K_{\star}}\right)
\eeq
Since the scale $t_\star$ is unphysical, one can then resum these logs by the usual RG treatment by demanding that the bare correlators are independent of $t_\star$.

Unfortunately, the scheme defined here does not work for metric fluctuations directly (although it can be used for additional field during inflation). Diffeomorphism invariance does not allow us to change the time-dependence or scaling dimensions of $\zp$ and $\zm$ independently of the background. As a result, we must modify the above treatment. Fortunately, for many practical calculations, the tilt of the spectrum provides the natural regulator we need for an RG treatment. This can be easily demonstrated in the case of mixing of composite operators, where by Wick contraction we have~\cite{Cohen:2020php}
\bea
\langle \zp^n \ldots\rangle &\supset& \langle \zp^{n-2} \ldots \rangle \times  \binom{n}{2} \int \frac{d^3 k}{(2\pi)^3} \frac{H(t_\star)^4}{4 \Mpl^2 \dot H(t_\star) c_s(t_\star)} \frac{1}{k^3} \left(\frac{k_\star}{k}\right)^{2 \epsilon + \eta + s} \\
&=& \langle \zp^{n-2} \ldots \rangle\binom{n}{2} \frac{H(t_\star)^4} {8 \pi^2 \Mpl^2 \dot H(t_\star) c_s(t_\star)}\left(- \frac{1}{n_s-1} -\gamma_E + \log \frac{K}{k_\star} +\ldots \right)
\eea
where $K$ is an IR regulator. Suppose that $k_\star \sim K$ so that log is already small and does not need to be renormalized. We still need to remove the $n_s-1$ pole, so we define the renormalized (composite) operator 
\beq
[\zp^n]_R = [\zp^n] + [\zp^{n-2}]\binom{n}{2} \frac{H(t_\star)^4} {8 \pi^2 \Mpl^2 \dot H(t_\star) c_s(t_\star)}\frac{1}{n_s-1} \ .
\eeq
Now we recall that $t_\star$ is an unphysical scale, and therefore the original correlators should not depend on it. Therefore, the renormalized operator obeys
\beq
\frac{d}{dt_\star} [\zp^n]_R = - [\zp^{n-2}]_R \binom{n}{2} \frac{H(t_\star)^4} {8 \pi^2 \Mpl^2 \dot H(t_\star) c_s(t_\star)}
\eeq
where we see that derivatives with respect to $t_\star$ cancel the $1/(n_s-1)$ to give the same result as we would for a massless scalar.

The possible mixing of composite operators of $\zeta$ is constrained by self-consistency with the nonlinearly realized symmetries of $\zeta$~\cite{Cohen:2021jbo}. In addition, all these equations can be recast, using integration by parts, as single RG equation for the probability distribution of $\zp$. The result is that
\beq
\frac{\partial}{\partial \t} P(\zp, \t)=\sum_{n \geq 2}(-1)^n \frac{\gamma_n}{n!} \frac{\partial^n}{\partial \zp^n} P(\zp, \t) \ ,
\eeq
where $\t = H t$ and we used the equations of motion to relate $t_\star$ and $t$, as is familiar from the Callan–Symanzik equation. Intuitively, the nonlinear dilation, $\zp \to \zp+ \lambda$, only allows $\zp$ to appear in derivatives. The coefficients $\gamma_n$ are determined by the mixing of $[\zp^n] \to 1$.

This equation is important because it allows us to determine the allowed renormalization group evolution of $\exp(q \zp)$ which appears throughout our action. Specifically, we can use integration by parts to write
\bea
\frac{d}{d\t} \langle \exp(q \zp) \rangle  &=& \int d\zp \exp(q\zp) \frac{\partial}{\partial \t} P(\zp, \t)\\
&=& \int d\zp \exp(q\zp)\sum_{n \geq 2}(-1)^n \frac{\gamma_n}{n!} \frac{\partial^n}{\partial \zp^n} P(\zp, \t) \\
&=& \langle \exp(q \zp) \rangle  \sum_{n \geq 2} \frac{q^n \gamma_n}{n!} 
\eea
 We see that $\exp(q\zp)$ does not mix with other operators and therefore can be renormalized by a single counter-term
\beq
[\exp(q \zp )]_R = Z_q [\exp(q \zp )]
\eeq
This result can also be proved directly at the operator level, but is more straightforward and intuitive in terms of the Stochastic description. The result is important because it means that all the IR divergences associated with contractions of $\zp$ that arise from expanding $\exp(q\zp)$ must recombine into an exponential. Moreover, when the theory is under perturbative control, the coefficients $\gamma_{n} \propto \Delta_\zeta^{n/2} \ll 1$ and thus does not meaningfully alter the power counting of the EFT. 

The special behavior of $e^{q \zp}$ under renormalization has a nice physical interpretation in terms of the operator for the volume of the universe at the end of inflation. Specifically, it is natural to define a family of (renormalized) volume operators as
\beq
\hat V^n = L^3 a^{3n}(t) e^{3 n \zp} \quad \to \quad [\hat V^n]_R = L^3 a^{3n}(t)[\exp(3n \zp )]_R \ ,
\eeq
where $L$ is some reference comoving scale (often taken to be the size of the initial inflationary patch). These operators play an important role in characterizing the onset of eternal inflation. Specifically, $\hat V$ was defined in~\cite{Creminelli:2008es} as an order parameter for the phase transition, while statistics of $[\hat V^n]_R$ characterizes the full probability distribution~\cite{Dubovsky:2008rf,Dubovsky:2011uy,Lewandowski:2013aka,Cohen:2021jbo}. The appearance of these exponential operators in the volume is required since a constant value of $\zeta=\lambda$ is equivalent to a change of the scale factor $a(t)\to e^\lambda a(t)$. 

\section{All-Orders Conservation}\label{sec:all_orders}

Conservation of $\zp$ to all orders is physically intuitive classically, for example via the separate universes description of the long wavelength modes. Since the long wavelength mode is locally indistinguishable from an unperturbed FRW background. Time evolution of $\zeta$ would seem to break this description. However, the coupling of quantum (short-wavelength) fluctuations to the long modes complicates this simple picture and has required complicated diagrammatic arguments to understand. Our goal in this section it to explain how the conservation of $\zeta$ survives these quantum corrections in a way that is simple and intuitive, yet is also manifest in (some) methods for performing the loop calculations.

To start, we will use the results from the EFT description to show that power counting and symmetries ensures conservation of the super-horizon $\zp$ operator, even when coupled to additional massive degrees of freedom. We will then show how explicit loop calculations in Mellin variables realize this power counting behavior in the full UV description.

\subsection{Conservation from Power Counting}

In the previous section, we showed that within the EFT of soft $\zeta$ in the slow-roll regime, the action can be written in the form
\bea
S_{\rm EFT} &=& \int d^4x e^{3 \zp} \bigg( \kappa\dot \rho \dzp \zm - \kappa(t)\rho(t) \Lambda(t)^{-2} e^{-2 \zp}\partial_i \zp \partial^i \zm \\ &&+c_n(t_\star) (\Lambda(t))^{-n +\gamma_n(t_\star)} (e^{-\zp} \partial \zp)^n \zm + \ldots \bigg) \ .
\eea
where $\gamma_n(t_\star) \ll 1$ is the anomalous dimension which includes renormalization of $c_n(t_\star)$ from $\exp(q\zp)$ and slow-roll suppressed terms. The $\ldots$ includes terms with more derivatives per operator, non-local terms and higher powers of $\zm$. As we have demonstrated, all such terms are suppressed by at least one power of $\Lambda(t)$ and thus are formally irrelevant.

The irrelevance of interactions of at long wavelengths is not limited to single field inflation, but persists in couplings to additional fields, provided that their momentum space correlators all decay as a positive power\footnote{We do not include factors of $c_s(t)$ here as $c_s(t)$ may not be scale relevant to non-inflaton degrees of freedom. As we are working in the regime where the time dependence of any parameter is small in Hubble units, power law decay in $a(t)H(t)$ is sufficient.} of $k/(a(t)H(t))$. Specifically, we will suppose there is an operator $\O_\chi$ associated with some additional degrees of freedom $\chi$ that obeys
\beq
\langle \O_\chi(\k_1,t) ... \O_\chi(\k_n,t) \rangle) \propto (a(t) H(t))^{-n \Delta_\chi}
\eeq
when $k_i \ll aH$ for all $i=1,..,n$. In this case, we can treat $\Delta_\chi ={\cal O}(1) > 0$ is the effective dimension of the operator and rescale $\O_\chi = (a(t) H(t))^{-n \Delta_\chi}\tilde{\mathcal O}_\chi$. We can couple this to $\zp$ and $\zm$, for example, via
\beq
S_{\rm int} \supset \int d^3 x dt \left( e^{3\zeta} c_\chi(t) (a(t) H(t))^{3-\beta - \Delta_\chi} \zm \tilde{\mathcal O}_\chi + c_\chi'(t) e^{\zp} (a(t) H(t))^{1-\Delta_\chi} \partial^2 \zp  \tilde{\mathcal O}_\chi  \right) 
\eeq
The first term is irrelevant by power counting while the second appears to be relevant when $\Delta_\chi < 1$ (which can still be consistent with $\Delta_\chi = {\cal O}(1)$. However, the second term is redundant, as we encountered for other operators that do not include factors of $\zm$. Specifically, we use the field redefinition
\beq
\zm \to \zm + \kappa^{-1} \int^t dt' c_\chi'(t') e^{-3\zp} (a(t'
) H(t'))^{1-\Delta_\chi}\partial^2 (e^{\zp} \tilde{\mathcal O}_\chi)
\eeq
After using integration by parts several times, we see that this cancels $c_\chi'(t)$, up to higher higher order in $k$ (note that $\dot{  \tilde{\mathcal O}}_\chi = 0$ at leading order, by definition).

In matching to the UV theory, it is possible to find contributions from contact operators~\cite{Cohen:2024anu}. These arise from the need to match contact terms generated by loops in the UV theory. They appear much like a generic coupling to an additional sector 
\beq
\begin{aligned}
S_{\rm int} \supset \int d^3 x dt \bigg( & e^{3\zeta} c_\Delta(t) (a(t) H(t))^{3-\beta - \Delta} \zm \tilde{\mathcal O}_\Delta \\
& + c'_\Delta(t) e^{\zp} (a(t) H(t))^{\Delta-2} \partial^2 \zp  \tilde{\mathcal O}'_{3-\Delta}  \bigg)
\end{aligned}
\eeq
The important feature of these contact operators is that their power spectra vanish, but 
\beq
\langle\tilde{\mathcal O}_\Delta (\k)\tilde{\mathcal O}'_{3-\Delta}(\kp) \rangle \propto \delta(\k+\kp)
\eeq
Because they do not have their own power spectra, they must appear in pairs and thus effectively behave as operators with $\Delta =3/2$ and are therefore irrelevant. It remains an interesting question as to whether these contact operators appear in single field inflation via matching, which we will address in the next subsection.

At this stage, we would argue that the conservation of $\zp$ is guaranteed by power counting. Since every interaction in the EFT decays with powers of $a(t) H(t)$, by the usual EFT rules, we would expect all perturbative corrections decay like $(k/(a H))$ and thus insure that the time independent initial conditions dominate the late time behavior. However, the observational that corrections come with inverse powers of $a(t)$ is not new and was not previously sufficient to constitute a demonstration of all-orders conservation. For example, in the all-order derivation of the conservation of $\zeta$ in~\cite{Assassi:2012et}, the vanishing of the commutator outside of the horizon was used to show that the long wavelength mode in single field inflation must obey an operator equation
\beq\label{eq:time_evol}
\dot \zeta=  \sum_n c_n (a^{-2} \partial^2 \zeta)^n + \ldots  \ , 
\eeq
where $\ldots$ includes operators with more and/or different distributions of derivatives. The list of operators on the right hand side (RHS) of this expression are very similar to the operators that multiply $\zm$ in the EFT action, as their form is constrained by the same nonlinearly realized symmetries. 

In~\cite{Assassi:2012et}, a diagrammatic proof was required to show that the RHS of Equation~(\ref{eq:time_evol}) vanishes as a powers of $k/(aH)$. The reason that this isn't self evident is that the fields that appear hear are the full UV operators. The UV description contains a scale $H(t)$ that can be defined by horizon crossing in comoving coordinates, $k = a(t) H(t)$. This scale lies in the middle of range of loop integration and thus an integral over internal momentum $p$ may generate a power of $aH$ rather than $k$. In contrast, the EFT is constructed so that all loop corrections are power-laws that can be regulated by analytic continuation the tilt or dimension of the operators. Furthermore, the scale $a(t) H(t)$ appears only as the UV scale that defines our couplings but does not appear as a scale in loop corrections.

\subsection{One-Loop Matching with Mellin}

The EFT argument for the conservation of $\zeta$ crucially relies on the existence of scaleless regulators. This is not a physical requirement, of course, but makes the power counting results clear and avoids the breaking of symmetries common to other regulators used in dS. Like any EFT, will still need to match the theory to the UV and this is most easily done if the UV theory can be regulated in the same way. We will see that this naturally leads us to introduce the Mellin representation to regulate the loops.

Let us take a concrete one loop example. We will consider quantum corrections to the classical power spectrum in single field inflation,
\beq
	\begin{aligned}
	\langle \zeta (\k) \zeta(\kp) \rangle &= i^2 \int d^4x \, [H_{\rm int}, [H_{\rm int}, \zeta(\k) \zeta(\kp)]] \ .
	\end{aligned}
\eeq
We will consider the case where $H_{\rm int} = -\lambda a^3 H^{-3} \dot{\zeta}_c^3$, $\zeta_c$ is the dimensionless canonically\footnote{Technically speaking, we have absorbed an extra factor of $H \sqrt{2}$ into the definition of $\zeta_c$ so that $\Delta_\zeta$ is the amplitude of the power spectrum and $\zeta_c$ is normalized to 1. } normalized field such that $\zeta = \Delta_\zeta \zeta_c $
and $c_s = 1$. The coupling $\lambda$ can arise, for example, from
\beq
{\cal L}_{\rm int} \supset \frac{4 M_3^4}{3} \frac{ \dot \zeta^3 }{ H^3} \quad \to \quad \lambda = \frac{8 \sqrt{2} M_3^4}{3 H^3 }\Delta_\zeta^3 \ ,
\eeq
in the EFT of inflation. This example displays all the relevant challenges, but we calculate additionally the contributions from other interactions in the Appendices. 

To proceed, we define the canonically normalized $\zeta_c$ operator as
\begin{equation}
	\zeta_c = \int \frac{d^3k}{(2\pi)^3} \left (a_{\k} \bar \zeta_c(\k,\tau) + a_{-\k}^{\dagger} \bar \zeta_c^* \right )
\end{equation}
where 
\begin{equation}
	\bar \zeta_c = -i\sqrt{\frac{\pi}{2}} e^{\frac{1}{2}i \pi(\nu + 1/2)} \tau^{\nu} {\rm H}^{(1)}_\nu \left( - k \tau \right) \ ,
\end{equation}
and $2\nu= 3 + 2\epsilon + \kappa  = 3 - (n_s-1)$(remembering that since $c_s=1$, we have $s=0$). We then insert this into the Wick contracted expression
\begin{align}
\langle \zeta(\k) \zeta(\kp) \rangle =& -\lambda^2 \int_{-\infty}^{\tau} d\tau_1 a^4(\tau_1) \int_{-\infty}^{\tau_1} d\tau_2 a^4(\tau_2)\int \frac{d^3k} {(2\pi)^3} \times \\
	 & \bar \zeta(\k, \tau) \bar \zeta(\kp, \tau) \dot{\bar{\zeta}}^* (\k, \tau_1) \dot{\bar{\zeta}}^* (\kp, \tau_2) \dot{\bar{\zeta}} (\p, \tau_1) \dot{\bar{\zeta}}^* (\p, \tau_2) \dot{\bar{\zeta}}(\p - \k, \tau_1) \dot{\bar{\zeta}}^* (\k - \p, \tau_2) \nonumber \\
  &\qquad+ \text{3 other terms} \nonumber
\end{align}
Here we are dropping the other terms associated with the two commutators, but the complete details can be found in the Appendix.

Performing loop calculations directly with these Hankel functions makes regulating the theory very difficult. We will therefore use the Mellin representation,
\begin{equation}
	i \pi e^{i\pi \nu/2} H^{(1)}_{\nu} (z) = \int^{c+i\infty}_{c-i\infty} \frac{ds}{2\pi i} \Gamma \left ( s + \frac{\nu}{2} \right ) \Gamma \left ( s - \frac{\nu}{2} \right )\left ( -\frac{i z}{2} \right )^{-2s} \ ,
\end{equation}
where $c$ is a real number chosen so that we can enclose all the poles of the $\Gamma$-functions. Since $\nu$ is real, we can choose $c>\nu/2$ and close the contour in the ${\rm Re}s<0$ region to isolate both sets poles at $s=\pm\nu/2-n$, where $n=0,1,...$ is an integer, with residues $(-1)^n/n!$. Using the residue theorem, one recovers the series expansion of the Hankel function around $z=0$. 

While the full calculation is quite involved (see Appendix~\ref{app:mellin} for complete details), but the main result can be understood from a few simple observations. First, the time derivatives act on the Mellin transformed mode functions as
\bea
\dot{\bar \zeta}_c &=& C \int^{c+i\infty}_{c-i\infty} \frac{ds}{2\pi i} (\nu-2s) \Gamma \left ( s + \frac{\nu}{2} \right ) \Gamma \left ( s - \frac{\nu}{2} \right )\left ( \frac{i k \tau}{2} \right )^{-2s} (-\tau)^{\nu} \\
&=&  C \int^{c+i\infty}_{c-i\infty} \frac{ds}{2\pi i} (-2) \Gamma \left ( s + \frac{\nu}{2} \right ) \Gamma \left ( s - \frac{\nu}{2} +1\right )\left ( \frac{i k\tau }{2} \right )^{-2s} (-\tau)^{\nu}  \ .
\eea
On the second line, we see that the poles of the second $\Gamma$-function have been shifted to $s= \nu/2 - (n+1)$ for $n=0,..,\infty$. This is expected, since in the $k \tau \to 0$ limit, there is always constant solution whose time derivative is zero. This is important because after loop integration divergences will be associated with coincident poles of different $\Gamma$-functions, so this is a this results is more meaningful than simply what it reveals about the series expansion.

The second important feature of the Mellin transform is that it dramatically simplifies loop and time integrations. After inserting the Mellin transform for each mode function, we will be left with a loop integral
\beq
\int \frac{d^3 p}{(2\pi)^3} p^{-2 s_3 -2 s_4} (|\k-\p|)^{-2 s_5-2 s_6}  = -\frac{i}{2\pi} \delta \left( \frac{3}{2} - (s_3 + s_4 + s_5 + s_6 )\right) + {\cal O}(k^2) 
\eeq
where $s_{i=3,4,5,6}$ are the Mellin variables that define the Hankel functions for the four Hankel functions in the loop integral and $s_{1,2}$ will be the Mellin variables for the fields carrying momentum $\k$ and $\kp$ contracted with the external operators. On the right hand side, we have written the solution as an expansion in $k$. It is straightforward to check that the original integral is IR finite and therefore we can treat the integral as an expansion in $k/p$. 

The time integrals are similarly straightforward. Prior to evaluating the Mellin integrals, we have only integrals over power laws in $\tau$ and $\tau'$ to evaluate. Importantly, unlike the loop integrals, the power-law depends on $s_1$ and $s_2$. To simplify the discussion, we will drop slow roll corrections, so that $\beta \to 3$ and $a(\tau) \propto \tau^{-1}$. In this limit, we will encounter integrals of the form
\begin{align}
\int_{-\infty}^{\tau} d \tau_1 \,(\tau_1)^{\frac{1}{2}} (\tau_1)^{-2(s_1 + s_3 + s_5)} &= \frac{\tau^{\frac{3}{2} -2(s_1+s_3+s_5)}}{\frac{3}{2} -2(s_1+s_3+s_5)} \nonumber \\
&\to \frac{i \pi}{2} \delta \left ( \frac{3}{2} - 2(s_1 + s_3+s_5)  \right ) \ ,
\end{align}
or of the nested form
\begin{align}
\int_{-\infty}^{\tau} d \tau_1  & (\tau_1)^{-2(s_1 + s_3 + s_5)}\int_{-\infty}^{\tau_1} d \tau_2(\tau_2)^{\frac{1}{2}} (\tau_2)^{-2(s_2 + s_4 + s_6)}  \\
&\to \frac{1}{\frac{3}{2} -2(s_2+s_4+s_6)} i \frac{\pi}{2} \delta(3 - 2(s_1+\ldots s_6)) \ .
\end{align}
In order to get time-independent corrections, we need to saturate the $\delta$-functions associated with these time-integrals as well as the $\delta$ function from the loop momentum integral.

Before taking the $\tau \to 0$ limit, the integrand will include the following factor
\bea
&& \frac{(-\tau k)^{-2s_1-2s_2} }{k^{3 - (n_s-1)}}  \tau^{3-2(s_3+s_4+s_5+s_6)} \delta \left( \frac{3}{2} - s_3 + s_4 + s_5 + s_6 \right) \nonumber \\
&&=   \frac{(-\tau k)^{-2s_1-2s_2} }{k^{3 - (n_s-1)}} \delta \left( \frac{3}{2} - s_3 + s_4 + s_5 + s_6 \right)\ ,
\eea
where the $\delta$-function is from the loop integration. This form does not depend on the nature of the time integral. Therefore, any correction survives the $\tau \to 0$ limit must be consistent with $s_1 + s_2 = 0$.

Given the form of the integrand, it remains possible to generate a logarithmic correction, $\log k \tau$, if the poles of the $\Gamma$ functions coincide when $s_1=-s_2$. This condition arises naturally from the poles associated with the time integration. Substituting this condition, by performing the $s_2$ Mellin-integral for example, the integrand of the $s_1$ Mellin integral becomes
\beq
\langle \zeta^2 \rangle_{\rm 1-loop} \propto \int \frac{ds_1}{(2\pi i)} \Gamma \left ( s_1 + \frac{\nu}{2} \right ) \Gamma \left ( s_1- \frac{\nu}{2} +1\right )\Gamma \left ( -s_1 + \frac{\nu}{2} \right ) \Gamma \left ( -s_1 - \frac{\nu}{2} +1\right ) \ .
\eeq
The poles of the integrand a located at $s_1 = \pm \nu/2 \pm n $ and $s_1 = \mp (\nu/2-1) \pm n$ where $n=0,1,2,...$. Since $\nu / 2$ is not an integer, these poles can only coincide if 
\beq
\nu/2  + n =  (\nu/2-1) - n'  \quad \to \quad n+n' =-1 \ ,
\eeq
where $n,n' \in 0,1,2,...$. Clearly this condition is never satisfied and therefore no logarithmic correction arises.

\subsection{Scheme Dependence}

The purpose of the Mellin calculation was to demonstrate, unambiguously, that we can regulate the 1-loop correction to the power sprectrum without any signs of divergences or logarithmic secular terms. The purpose of our soft $\zeta$ EFT is to explain this behavior physically. However, it does not explain why this was no apparent in other approaches to calculating the one-loop power spectrum.

As a concrete illustration, we can consider the calculation of~\cite{Senatore:2009cf}, which found for $-\frac{4}{3} M_3^4 \dot \zeta^3$ the correction to the power spectrum is given by
\beq
\left\langle\zeta_{\vec{k}}(t) \zeta_{\vec{k}^{\prime}}(t)\right\rangle'_{1-\text { loop }}=- \frac{2}{15 \pi^2} \frac{ H^8 M_3^8}{\dot{H}^4 M_{\mathrm{Pl}}^8} \frac{1}{k^3} \log \left(\frac{H}{\mu}\right) \ .
\eeq
However, it useful to work in a scheme where $\mu = H$ and this result vanishes~\footnote{It should be emphasize that the appearance of $\log \mu /H$ is itself a non-trivial property of the full 1-loop calculation in many schemes. Earlier attempts to calculate this result had found $\log k$ which would have broken important symmetries of single field inflation~\cite{Weinberg:2005vy,Chaicherdsakul:2006ui,Seery:2007we,Dimastrogiovanni:2008af,Adshead:2008gk}. This should emphasize the value of regulators, like the EFT or Mellin, where the absence of such logs is manifest.}. This is effectively the scheme that Mellin implements automatically, as one does not need the introduction of a scale $\mu$, as all scales are defined in terms of $H(t_\star)$. In other schemes, this log may appear as a large constant~\cite{Kristiano:2022zpn} that would diverge as $n_s \to 1$. If such a term was universal, it would appear in Mellin space as the collision of poles, but we have verified explicitly that does not occur in our Mellin-based scheme.

A source of the confusion is that previous calculations have been defined in dimensional regularization, which is supposed to handle these problems well. However, when calculating in-in correlators, it is often advantageous to treat time and space separately. Since the theory is invariant under spacial translations, we can Fourier transform the spacial coordinates $\x \to \k$. However, specifying a unique time that defines an equal time correlator breaks time translations, making it less useful to Fourier transform. 

In flat space, loop corrections to in-in correlators using this decomposition would require performing integrals of the form
\beq
\int \frac{d^3p}{(2\pi)^3} p^n e^{ i m p t} \ ,
\eeq
where the factor $e^{i m p t}$ is from the time evolution of the mode functions. Notice that if $t=0$, this is a scaleless integral with a powerlaw divergence and would vanish,
\beq
\int \frac{d^3p}{(2\pi)^3} p^n = 0 \ .
\eeq
However, at $t \neq 0$, the integral is no longer scaleless and we find a non-zero answer (by continuing $t \to t (t- i \epsilon)$ with $t<0$),
\beq
    \int \frac{d^3p}{(2\pi)^3} p^n e^{ i m p t - \epsilon m p |t| } \to \frac{1}{2\pi^2 }\frac{\Gamma[3+n]}{(-i mt)^{3+n}}\ .
\eeq
Importantly, this expression diverges as $t \to 0$, which demonstrates that the power law divergence at $t=0$ is not actually regulated in this scheme.

Of course, power law divergences should be removed in any scheme by a redefinition of the couplings, without introducing any need for RG flow. Therefore, even though the bad behavior appears to be time dependent, it should be removable by a change to the action. Indeed, it was argued in~\cite{Senatore:2009cf} that the problematic behavior associated with $\dot \zeta^3$ can be removed by a shift of the coupling of a $\dddot \zeta^2$ operator. In Appendix~\ref{app:flat}, we show this same behavior can be seen when computing a flat space in-in correlator with the same interactions. In this precise sense, these different calculations differ only in their effective values of the couplings of the theory as a result of the choice of regularization scheme.

However, the Mellin-space approach is not free from the challenge of defining a simple scheme. If we follow the conventional set of contours for the remaining Mellin integrals integrals, we will arrive at a finite correction to the power spectrum
\beq
\langle \zeta(\k) \zeta(-\k)\rangle' =  \frac{\Delta_\zeta k^{n_s-1}} {k^3}\left(1+ a \lambda^2 \right) \ , 
\eeq
where $a$ is a finite constant. In Appendix~\ref{app:mellin}, we calculate $a$ for the $\lambda \dot \zeta^3$ interaction, and find $a =-9.36 \times 10^4$ for a particular choice of contours (for one subset of the diagrams). However, as described in the Appendix, there is another choice of contours which is equally physical that gives rise to $a=0$ for all the 1-loop Mellin corrections. This should be understood as the statement that this one-loop correction is equivalent to the shift of $\dddot \zeta^3$ coupling and therefore we can choose a scheme where the one-loop correction is zero, and any finite correction is thus determines by this additional operator. In other schemes, the constant $a$ may even be infinite (because the divergence is not actually regulated), and should be removed by a counter-term by hand as usual. 

The issue of how to apply a self-consistent scheme in Mellin space is challenging beyond one-loop~\cite{Premkumar:2021mlz}. The collision between poles remains a physical effect associated with RG. However, after regulating the theory, the choice of contours will shift the final results by a finite amount. Since these choices do not change the $k$ or $\tau$ dependence of the final result, they are consistent with shifts of local couplings. On the other hand, the proliferation of Mellin variables introduces a significant challenge to regulating and renormalizing the theory at higher-loop order.

\section{EFT with Tensor Modes} \label{sec:tensors}

So far, we have focused only on the scalar metric fluctuation, $\zeta$. This might seem surprising, as the tensor modes, $\gamma_{ij}$, are present in both inflation and de Sitter, and are the more basic ingredient of dynamical gravity. However, in practice, the scalars and tensors have essentially the same behavior in perturbation theory\footnote{The observational signatures of tensor modes~\cite{Seljak:1996gy,Kamionkowski:1996ks} and their statistics~\cite{Maldacena:2011nz,Meerburg:2016ecv,Dimastrogiovanni:2018uqy,Bordin:2020eui,Cabass:2021iii,Cabass:2021fnw,Cabass:2022jda} are importantly very different from the scalar fluctuations. These differences only indirectly impact the long wavelength theory, but not the structure of the EFT. }, except that tensor structures and symmetries lead to additional constraints~\cite{Creminelli:2014wna,Bordin:2016ruc}. Our goal in this section is to demonstrate that EFT and conservation for the the soft tensor and scalar modes are effectively identical.

\subsection{Tensors from the Top Down}

The tensor modes, defined via $g_{ij} = a^2(t) e^{2 \zeta} [e^{2 \gamma}]_{ij}$, can also be treated through the EFT techniques. The action for the tensors is given by
\begin{equation}
	S = \frac{\Mpl^2}{8}\int d^4x\, a^3(t) \big[ \dot{\gamma}^2_{ij} - \frac{1}{a^2} (\partial_i \gamma)^2\big]
\end{equation}
Solving for the time dependence, we can decompose the tensor modes into the growing and decaying modes as well
\begin{equation}
	\gamma_{ij} = \gamma_{+, ij} + (aH)^{-\beta} \gamma_{-, ij}
\end{equation}
where we have that $\beta = 3 + 2 \epsilon$ for the case of the tensors. We will have 
\begin{equation}
	\dot \gamma = \dot{\gamma}_{ij} - \beta H(t)(1-\epsilon) (a(t) H(t))^{-\beta} \gamma_{-, ij} + (a(t) H(t))^{-\beta} \dot{\gamma}_{-,ij}
\end{equation}
Then, we can go through the same exercise, and we will find that 
\begin{equation}
	S = \frac{\Mpl^2}{4}\int d^4x\, \big[ - \beta H \dot{\gamma}_{+, ij} \gamma_{-, ij} + \dot{\gamma}_{+, ij} \dot{\gamma}_{-, ij} - \frac{1}{a^2} \partial_i \gamma_{+, ij} \partial_i \gamma_{-, ij} \big]
\end{equation}
to lowest order in slow roll parameters. Again, the canonical momenta is given by 
\begin{equation}
	\Pi_{\gamma_+} = -\frac{\Mpl^2}{4}\beta H \gamma_-
\end{equation}
with the relation 
\begin{equation}
\begin{aligned}
    [\gamma_{+,ij}(\x,t),\gamma_{-,lm}(\x',t)] &= -4\frac{i }{\Mpl^2 \beta H} \delta(\x-\x') \delta_{il} \delta_{jm} \\
    \to[\gamma_{+,ij}(\k,t),\gamma_{-,lm}(\kp,t)] &= -\frac{4i}{\Mpl^2 \beta H} \delta_{il} \delta_{jm} (2\pi)^3 \delta(\k+\kp) 
\end{aligned}
\end{equation}
Much of the story of $\zeta$ follows through into the tensors. The canonical momenta relations enforce that the leading order interactions will be of the form $\supset \gamma_+^n \gamma_-$. Moreover, as we saw in the case of $\zeta$, $[e^{\gamma}]_{ij}$ forms a nice operator which renormalizes into itself under RG. The only subtlety arises when we try to understand the operators that are allowed under symmetry transformations for $\gamma_{ij}$, since $\gamma_{ij}$ has its own symmetry transformations. This is required to conclusively show that any operator that arises in the theory is irrelevant. In particular, we would want to show that no non-local terms can give rise to operators which become relevant outside the horizon, as we did with $\zeta$. Let us turn to that next.

\subsection{Symmetries of scalar and tensors}

The tensor symmetries are different from the scalar symmetries~\cite{Hinterbichler:2012nm}. Tensors change linearly under dilatation, 
\begin{equation}
	\delta \gamma_{ij} = \lambda x^k \partial_k \gamma_{ij} \ ,
\end{equation}
while the SCTs are not a good symmetry of the tensors. Instead, tensors have their own set of non-linearly realized symmetries. Under a general gauge transformation, $x^i \to x^i + \xi^i$, the tensors transform according to
\begin{equation}
 	\delta \gamma_{ij} = \partial_i \xi_j + \partial_j \xi_i
\end{equation} 
for time independent $\xi_i$. In terms of the soft modes, this breaks up as
\begin{equation}
	\delta \gamma_{+, ij} =  \partial_i \xi_j + \partial_j \xi_i \qquad \delta \gamma_{-, ij} = 0 \ .
\end{equation}
Although any $\xi_i$ represents a valid diffeomorphism, only those that leave the gauge unchanged and are continuously connected to physical modes are of interest for understand the EFT. Concretely, for the our choice of gauge, we will only include the transformations which keep the tensors transverse and traceless. This construction is outlined in \cite{Hinterbichler:2013dpa}.

Now we want to understand the structural constraints imposed on $\gamma$ and $\zeta$ when taking into account the tensor symmetries. Instead of taking a dilatation, let us take a gauge transformation $x^i \to x^i + \xi^i$, where $\xi_i = M_i^{\,\,j} x_j$, where $M_{ij} = M_{ji}$ is a constant symmetric matrix. We can then capture the non-linear shifts in the field as
\begin{equation}
    \delta \zp = \frac{1}{3} \partial_i \xi_i \qquad \delta \gamma_{+, ij} = \partial_i \xi_j + \partial_j \xi_i
\end{equation}
The corrections are of higher order in $\zp, \gamma_{+}$. Physically, $\delta \zp$ captures the dilatation, while $\delta \gamma_{+}$ captures anistrotropic scaling. The charge corresponding to this symmetry is given in terms of the operator
\begin{equation}
	Q = \int d^3x\, \partial_i \xi_j \left ( \frac{1}{3} \delta^{ij} \Pi_{\zp} + 2 \Pi^{ij}_{\gamma_+} \right ) = \int d^3x\, M_{ij} \left ( \frac{1}{3} \delta^{ij} \Pi_{\zp} + 2 \Pi^{ij}_{\gamma_+} \right )
\end{equation}
Fourier transforming, and using $\zm, \gamma_{-}$ as conjugate momenta, we have
\begin{equation}
	Q = \int d^3q \, \left ( \frac{C_1}{3} M_{ii} \zm (\q) + 2 C_2 M_{ij} \gamma_-^{ij}(\q) \right )
\end{equation}
Now, playing the same game as with $\zp$, we can demand that the action is invariant under transformations generated by the charge by definition. We will then have that
\begin{equation}
S\supset \int d^4 x \sqrt{g} \sum_n { \cal O}_n(x) \to [Q,\int d^4 x \sqrt{g} \sum_n {\cal O}_n(x)] = 0
\end{equation}
Any potentially marginal operator, ${\cal O}_m$, will take that form ${\cal O}_m = \zm F(\zp, \gamma_{+, ij}, \partial_t, \vec \partial, )$ or ${\cal O}'_m = \gamma_{-, ij} F^{ij}(\zp, \gamma_{+, ij}, \partial_t, \vec \partial, )$ where $F(\zp,\gamma_{+, ij}, \partial_t, \vec \partial), F^{ij}(\zp,\gamma_{+, ij}, \partial_t, \vec \partial)$ may be non-local in $\vec \partial$. Without loss of generality, we can take the former case. Now we can Fourier transform the fields so that 
\beq
\begin{aligned}
	\int d^4x \sqrt{g} {\cal O}_m \to  c_m(t) \zm(\p_0) &\left( \prod_{i=1}^N \int d^3 p_i \zp(\p_i)  \prod_{j=1}^N \int d^3 q_j \gamma_+(\q_j) \right) \\ &\times \tilde F(\{ p_i, q_j \} ) (2\pi)^3 \delta(\sum_{n=0}^n \p_{n} + \q_{n}) \ .
\end{aligned}
\eeq
Since the shift symmetry arises from $Q= \alpha \zm(\q=0) + \beta_{ij} \gamma_{-, ij} + {\cal O}(\zp \zm)$, a single commutator will set one of the momenta of $\zp(\p_i)$ and $\gamma_{+, ij}(\q_i)$to zero, 
\begin{equation}
[Q, \zp(\p)] \to \alpha i (2\pi)^3 \delta(\p) +{\cal O}(\zp) \qquad [Q, \gamma_{+, ij}(\q)] \to \beta_{ij} i (2\pi)^3 \delta(\q) +{\cal O}(\gamma_+) \ . 
\end{equation}
Therefore, after $N$ commutators we have 
\begin{equation}
[Q,[Q,...[Q,S],...,],] = 0 \supset \zm(\p_0=0) F(\{\p_i, q_i \} = 0 ) + {\cal O}(\zm \zp)   \ .
\end{equation}
Now, in order for the charge to be a symmetry, we need this to vanish in the limit that $p_i, q_j = 0$. This then requires that there are more derivatives in the numerator than in the denominator, and hence the non-local terms must be power counting suppressed even when taking both $\zeta, \gamma$ into account. 

\subsection{All-Orders Conservation}

The all-orders conservation of both $\gamma_+$ and $\zeta_+$ is again a result of power counting. As every interaction must appear with a factor of $\zm$ or $\gamma_i$ as before so that we have
\begin{align}
&\int d^4x \sqrt{g} (c_m {\cal O}_m + c_m' {\cal O}'_m) \to \nonumber\\
&\int dt d^3 x \left( c_m (aH)^{3-\beta_\zeta -\Delta_F}  \zm F(\zp, \gamma_{+, ij} \zeta)  + c'_m (aH)^{3-\beta_\gamma -\Delta_{F_{ij}}}  \gamma_{-,ij} F^{ij}(\zp, \gamma_{+, ij} \zeta) \right)
\end{align}
where $\Delta_F$ and $\Delta_{F_{ij}}$ and the scaling dimensions of the composite operators $F$ and $F^{ij}$.  As we saw in the previous section, because there must be a net positive number of derivatives in order to maintain the symmetries of the theory, $\Delta_F , \Delta_{F_{ij}} \geq 1+ {\cal O}(\epsilon)$ and therefore all such operator are irrelevant. 

The all orders result then follows from the scheme for calculating scaleless integrals in the EFT, as we discussed in Section~\ref{sec:all_orders} and in~\cite{Cohen:2020php}. Here the tensor structure present no additional challenges beyond the scalar case, as the momentum integrals are effective the same. However, the additional complication is that the speed of propagation of the scalars and tensors are different, in general, and therefore the power counting can be in both $k/(a(t)H(t))$ and $k / \Lambda(t)$ (i.e.~differ by factors of $c_s(t)$). This has no impact on the conservation of $\zeta$ or $\gamma$, provided that the time variation of the parameters is small. However, when determining the evolution of composite operators or the tilt of the spectra, there is no long a single $t_\star$ that makes logs for both $\gamma$ and $\zeta$ small for the same wavenumber $\k$, and therefore matching and RG can require additional resumming of logs~\cite{Baumann:2014cja}.

Finally, the conservation of the tensor modes is more robust to the addition of light fields due to the Higuchi bound~\cite{Bordin:2016ruc}. Specifically, the addition of spin-2 states is highly constrained by unitarity (positivity of the two-point statistics), as is well known from flat space. Therefore, even in the presence of additional fields, any time evolution would take the form
\beq
\dot \gamma_{+,ij} = {\cal O}_{ij}
\eeq
would require a tensor operator ${\cal O}$ that is close to dimension zero. We have already seen this is not permitted for the metric fluctuation itself, and can only arise for partially massless tensor modes. As a result, even in generic multifield models where $\zeta$ can evolve (do to isocurvature modes), the tensors will be generically conserved.

\section{Conclusions}\label{sec:conclusions}

Understanding perturbative quantum gravity in accelerating spacetimes is an important step towards resolving the deepest questions about the quantum nature of our universe~\cite{Flauger:2022hie}. The breakdown of our understanding is most acute in cases like eternal inflation~\cite{Vilenkin:1983xq,Linde:1986fd}, where the fluctuations of the metric are large. However, we can learn a lot about these non-perturbative questions from (renormalization group improved) perturbative calculations~\cite{Arkani-Hamed:2007ryv}. Concretely, studies of both eternal inflation and the Hartle-Hawking wavefunction~\cite{Hartle:1983ai} have found close connections to the framework of stochastic inflation~\cite{Creminelli:2008es,Maldacena:2024uhs}, which is amenable to perturbative techniques.

In this paper, we have offered an EFT approach to understanding the all-orders behavior of both the scalar and tensor fluctuations during inflation. We showed that both are conserved in single and quasi-single field inflation due to the formal irrelevance of all interactions outside the horizon. The central reason for their conservation is that the metric is subject to an infinite group of nonlinearly realized symmetries that are the result of large diffeomorphisms. These symmetries do not permit the interactions and/or operator dimensions that would lead to super-horizon time evolution (provided we do not have large time dependence of the background).

The development of this EFT is part of a larger goal of understanding the nature of the metric fluctuations in accelerating universes non-perturbatively. Within the EFT description, RG flow can be translated into a Fokker-Plank equation~\cite{Starobinsky:1994bd}. At leading order, this is the framework of Stochastic Inflation, but the EFT allows for one to systematically improve this calculation by including higher derivative corrections. Yet, this approach also predicts the demise of the Stochastic framework for sufficiently rare fluctuations~\cite{Cohen:2022clv}. The question of how to calculate the tail of the distribution~\cite{Panagopoulos:2020sxp,Celoria:2021vjw,Creminelli:2024cge} has a variety of interesting implications both for observational implications, such as primordial black holes~\cite{Panagopoulos:2019ail}, and conceptually, in terms of the meaning of the wavefunction of the universe and the nature of eternal inflation. Putting our perturbative understanding of quantum gravity in these cosmologies on more straightforward foundation, both technically and conceptually, is hopefully a useful first step towards addressing these challenging problems.

\paragraph{Acknowledgements}
We are grateful to Dio Anninos, Prish Chakraborty, Clifford Cheung, Tim Cohen, Yiwen Huang, Austin Joyce, Akhil Premkumar, and Guanhao Sun for helpful discussions
We are supported by the US~Department of Energy under grant \mbox{DE-SC0009919}. 

\newpage
\appendix


\section{Deriving the EFT}\label{app:EFT}

The derivation of our soft EFT for the metric fluctuations requires a number of technical steps and matching to the full UV theory. In this appendix we explain some of these details that we bypassed in the main text.

\subsection{Stochastic Initial Conditions}

As emphasized in the main text, the initial conditions for $\zpm$ are not determined by the EFT and require matching to the full theory. We can derive the Gaussian initial condition from the quadratic action, 
\begin{equation}
	S = \frac{1}{2}\int d^4x \, \left(2 \frac{\Mpl^2 |\dot H|}{c_s^2 H^2(t)} a^3(t) \right) \bigg[ \dot{\zeta}^2 - c_s^2(\partial_i \zeta)^2 \bigg] \ .
\end{equation}
Assuming we can use the slow-roll expansion, the equations of motion for the mode functions are 
\begin{equation}
	\partial_\tau^2 \bar \zeta - \frac{1}{\tau}(1 + \epsilon)(2 + \eta - 2s)\partial_\tau \bar \zeta + c_s^2(\tau) k^2 \bar \zeta = 0 \ ,
\end{equation}
and can be solved directly to first order in slow-roll, ${\cal O}(\epsilon)$, to find 
\begin{equation}
	\bar \zeta = \Delta_\zeta \zeta_c = \Delta_\zeta A \tau^{(3 + 2\epsilon + \eta - 2s)/2} H^{(1)}_\nu \left (c_s(\tau) k \tau \right ) + \bigg( \text{c.c.} \bigg) \ ,
\end{equation}
where $H_\nu^{(1)}$ is the Hankel function of the first kind, and $\nu = (3 + 2 \epsilon + \kappa + s)/2$. To determine the normalization of the mode function, $A$, we take $k\tau \to \infty$ using
\begin{equation}
	H^{(1)}_\nu \left (c_s(\tau) k \tau \right ) \to \sqrt{\frac{2}{\pi c_s(\tau) k \tau}}  e^{i c_s k \tau - \frac{1}{4} i \pi (-3 + 2\nu)} \ .
\end{equation}
As a result, the canonically normalized field obeys
\begin{equation}
\begin{aligned}
	\zeta_c &\to 
	 - A \tau^{(2 + 2\epsilon + \eta - 2s)/2}\sqrt{\frac{2}{\pi c_s (\tau) k}} e^{i c_s (\tau) k \tau - \frac{1}{4} i \pi (-3 + 2\nu)} \ .
\end{aligned}
\end{equation}
We can now normalize to plane waves using the WKB approximation\footnote{We absorbed an extra factor of $\sqrt{2}$ from the unusual definition so that $\Delta_\zeta^2$ is the amplitude of the power spectrum.} so that 
\begin{equation}
	A = i\sqrt{\frac{\pi}{2}} e^{\frac{1}{4}i \pi(-3 + 2\nu)} = -i\sqrt{\frac{\pi}{2}} e^{\frac{1}{2}i \pi(\nu + 1/2)} \ .
\end{equation}
So overall, we have that
\begin{equation}
	\bar\zeta_c = -i\sqrt{\frac{\pi}{2}} e^{\frac{1}{2}i \pi(\nu + 1/2)} \tau^{(3 + 2 \epsilon + \eta  - 2s)/2} H^{(1)}_\nu ( c_s(\tau) k \tau) \ ,
\end{equation}
and the $\zeta$ operator is given by
\begin{equation}
	\zeta = \bar \Delta_\zeta \int \frac{d^3k}{(2\pi)^3} \left (a_k \zeta_c + a_k^{\dagger} \bar \zeta_c^* \right ) \qquad \bar \Delta_\zeta^2 = c_s^3 \Delta^2_\zeta = \frac{c_s^2 H^4}{4 \Mpl^2 |\dot H|} \ ,
\end{equation}
where $a_k, a_k^{\dagger}$ have the commutation relation, 
\begin{equation}
	[a_k, a_k^{\dagger}] = (2\pi)^3 \delta^3(k - k') \ .
\end{equation}
Now, in the soft limit $c_s k \tau \to 0$, we have
\begin{equation}
	H^{(1)}_\nu \left (c_s(\tau) k \tau \right ) \to \left ( -i(c_s(\tau) k\tau)^{-\nu} \frac{2^{\nu} \Gamma(\nu)}{\pi} + (c_s(\tau) k \tau)^{\nu} \frac{2^{-\nu} (1 + i \cot (\nu \pi)}{\Gamma(1 + \nu)} \right ) \ .
\end{equation}
Writing $\tau^{(3 + 2 \epsilon + \eta - 2s)/2} = \tau^{\nu - 3s/2}$ and $c_s(\tau) = c_{s,0} \tau^{-s}$, we have
\begin{equation}
\begin{aligned}
	\bar \zeta_c &\to 
	 e^{\frac{1}{2}i \pi(\nu + 1/2)} A + i e^{\frac{1}{2}i \pi(\nu + 1/2)} \tau^{2 \nu - 3s} B (1 + i \cot(\nu \pi)) \ ,
\end{aligned}
\end{equation}
where 
\begin{equation}
	\begin{aligned}
	 A = \frac{(c_{s,0} k)^{-\nu} 2^{\nu} \Gamma(\nu) }{\sqrt{2 \pi}} \qquad B = \sqrt{\frac{\pi}{2}} \frac{(c_{s,0}k)^{\nu} 2^{-\nu}}{\Gamma(1 + \nu)} \ .
	\end{aligned}
\end{equation}
As a result, in the limit $\k \to 0$, our fundamental field operator takes the form, 
\begin{equation}
\begin{aligned}
	\zeta &\to \bar \Delta_\zeta \int \frac{d^3k}{(2\pi)^3} \bigg ( A ( e^{\frac{1}{2}i \pi(\nu + 1/2)} a_k +  e^{-\frac{1}{2}i \pi(\nu + 1/2)}  a_k^{\dagger})\\
	 &+ iB (1 + i \cot(\nu \pi)) \tau^{2\nu - 3s}( e^{\frac{1}{2}i \pi(\nu + 1/2)}  a_k + e^{2i \nu \pi} e^{-\frac{1}{2}i \pi(\nu + 1/2)}  a_k^{\dagger}) \bigg ) \ .
\end{aligned}
\end{equation}
Since this is the operator in the long wavelength limit, it can be expressed in terms of the long wavelength fields, so that
\begin{equation}
	\zeta = \bar \Delta_\zeta\int \frac{d^3k}{(2\pi)^3} \bigg [ \bar{\zeta}_+ (e^{i \delta_\nu} a_k + e^{-i \delta_\nu} a_k^{\dagger} ) + i \tau^{2 \nu - 3s} \bar{\zeta}_- (e^{-i \delta_\nu} a_k - e^{i \delta_\nu} a_k^{\dagger} )  \bigg] \ , 
\end{equation}
where we have defined $\delta_\nu = \pi( \nu + 1/2)/2$, and we have
\begin{equation}
	\bar{\zeta}_+ = A\bar \Delta_\zeta = \bar \Delta_\zeta \frac{(c_{s,0} k)^{-\nu} 2^{\nu} \Gamma(\nu) }{\sqrt{2 \pi}} \hspace{1cm} \bar{\zeta}_-  = \bar \Delta_\zeta \frac{B}{\sin(\nu \pi)} =\bar \Delta_\zeta \sqrt{\frac{\pi}{2}}\frac{(c_{s,0} k)^{\nu} 2 ^{-\nu}}{\sin (\nu \pi) \Gamma(1 + \nu)}
\end{equation}
Now if we define
\begin{equation}
	\tilde{a}_\k = e^{i \delta_\nu} a_\k + e^{-i \delta_\nu} a_\k^{\dagger} \qquad \tilde{b}_\k = i (e^{-i \delta_\nu} a_k - e^{i \delta_\nu} a_k^{\dagger}) \ ,
\end{equation}
and $\zp = \bar{\zp} \tilde a_\k$ and $\zm = \bar{\zm} \tilde b_\k$, then power spectra become
\begin{align}
	\langle \zp \zp \rangle &= \bar \Delta_\zeta^2  \frac{(c_{s,0}k)^{-2\nu} 2^{2\nu} \Gamma^2(\nu) }{2\pi} (2\pi)^{3} \delta^3 (k + k') \\
 \langle \zm \zm \rangle &= \bar \Delta_\zeta^2 \frac{\pi}{2} \frac{(c_{s,0}k)^{2\nu} 2 ^{-2\nu}}{\sin^2 (\nu \pi) \Gamma^2(1 + \nu)} (2\pi)^{3} \delta^3 (k + k')  \ ,
\end{align}
with the tilt being $n_s - 1 = -(2\epsilon + \eta + s)$ which reproduces the well known results. Taking $\epsilon, \eta, s \to 0$ (and $\nu \to 3/2$) one recovers the expected result
\beq
\langle \zp \zp \rangle' = \frac{\bar \Delta_\zeta^2}{c_s^3 k^3} = \frac{\Delta_\zeta^2}{k^3} = \frac{H^4}{4 \Mpl^2 \dot H c_s} \frac{1}{k^3} \ .
\eeq
We could have arrived at this result in a more straightforward fashion by taking the slow-roll parameters to vanish from the outset, but the tilt plays a non-trivial role in loop corrections, and therefore we want to be able to derive this result from the most complicated expressions.

\subsection{Deriving the Soft Action from the Top Down}

In addition to matching the initial conditions, it is useful to understand the origin of the EFT directly from the structure of the long wavelength degrees of freedom. We therefore want to derive the soft action for $\zeta$ from the top down, following the same strategy as the derivation of SdSET. By construction, the long distance theory is defined in terms of
\begin{equation}
	\zeta = \zp + \left(\frac{aH}{c_s} \right)^{-\beta}\zm \ .
\end{equation}
Note here that $H = H(t)$ and $c_s(t)$ are time dependent. As discussed in the main text, symmetry enforces there is no time dependence associated with $\zp$. At this point $\beta$ is undetermined, but we find that $\beta$ is directly given by the tilt for the theory to be consistent. To streamline the derivation, we will assume a local quadratic action for $\zeta$ and substitute 
\begin{equation}
	\dot{\zeta} = \dzp - \beta (H - \epsilon H - s_c H) \left(\frac{aH}{c_s} \right)^{-\beta} \zm + \left(\frac{aH}{c_s} \right)^{-\beta} \dzm \ .
\end{equation}
into the quadratic action to find
\begin{equation}
	\begin{aligned}
	S &= \frac{1}{2}\int d^4x \, \left(2 \frac{\epsilon}{c_s^2} a^3(t) \right) \bigg[ \dot{\zeta}^2 - a^{-2}(\partial_i \zeta)^2 \bigg]\\
	&= \frac{1}{2}\int d^4x \, \left(2 \frac{\epsilon}{c_s^2} a^3(t) \right)\bigg[ \big(\dzp - \beta (H - \epsilon H - s_c H) \left(\frac{aH}{c_s} \right)^{-\beta} \\ & \qquad \qquad  +\left(\frac{aH}{c_s} \right)^{-\beta} \dzm \big)^2
    - a^{-2} \big(\partial_i (\zp + \left(\frac{aH}{c_s} \right)^{-\beta} \zm ) \big)^2 \bigg] \ .
	\end{aligned} 
\end{equation}
We have to remove all the unwanted degrees of freedom so we are left with an action for the physical degrees of freedom. We want to only write down terms that couple the two modes $\zp, \zm$ together. Repeating a similar set of field redefinitions discussed in the main text, we arrive at
\begin{equation}
	\begin{aligned}
	S &= \frac{\Mpl^2}{2} \int d^4x \, 2 \frac{\epsilon}{c_s^{2 - \beta}} a^{3-\beta} H^{-\beta} \bigg[ -2 \beta H (1 - \epsilon - s_c) \dzp \zm + 2\dzp \dzm -2 \frac{1}{a^2}\partial_i \zp \partial_i \zm \bigg]\\
	&= \int d^4x \, 2 \frac{\epsilon}{c_s^{2 - \beta}} a^{3 - \beta} H^{-\beta} \bigg[ - \beta H (1 - \epsilon - s_c) \dzp \zm + \dzp \dzm - \frac{1}{a^2} \partial_i \zp \partial_i \zm \bigg]  \ .
	\end{aligned}
\end{equation}
To further simplify the action, will define the time evolution of the parameters such that
\begin{equation}
    \epsilon = \epsilon(t^*) \left(\frac{aH(t)}{aH(t^*)}\right)^{\eta} \quad H(t) = H(t^*) \left(\frac{aH(t)}{aH(t^*)} \right)^{-\epsilon}  \quad c_s(t) = c_s(t^*) \left(\frac{aH(t)}{aH(t^*)} \right)^{s_c} \ .
\end{equation}
As a result, the time-dependence is
\begin{equation}
\begin{aligned}
	\epsilon(t) c_s^{-(2 - \beta)} a^{3 - \beta}(t) H^{1 - \beta}(t) &\to \epsilon(t^*) \left( \frac{a H}{a_* H_*} \right)^\eta c_s^{-(2- \beta)}(t^*) \left(\frac{aH}{a_*H_*} \right)^{s_c(\beta-2)}\\
    & a^{3 - \beta}(t) H^{1 - \beta}(t^*) \left( \frac{a H}{a_* H_*} \right)^{-\epsilon(1 - \beta)}\\
    & \to a^{3 - \beta + \eta + s_c (\beta - 2) - \epsilon(1 - \beta)}(t) f(t_*) \ ,
\end{aligned}
\end{equation}
where we are working to first order in the slow roll parameters in the action. We get that $\beta = 3 + 2\epsilon + \eta + s_c$, which is again the tilt. Here, we have defined $f(t_*)$ as 
\begin{equation}
\begin{aligned}
    f(t_*) = \epsilon(t_*) H(t_*)^{-2} (k_*)^{-(\eta + \kappa + 2 \epsilon)} c_s^{\beta - 2}(t_*) \ .
\end{aligned}
\end{equation}
Finally, the free action is given by
\begin{equation}
	S_2 = -2\int d^4x \, \beta \epsilon_* H_*^{-2} k_*^{(n_s - 1)} c_s^{\beta - 2} \bigg[ \dzp \zm \bigg] \ ,
\end{equation}
where $\epsilon_*, H_*$ are evaluated at $t_*$. The conjugate momenta becomes
\begin{equation}
	\Pi_+ = -2 \beta (H_*)^{-2} \epsilon_* (k_*)^{n_s - 1} c_s^{\beta - 2} \zm 
\end{equation}
and hence we have that
\begin{equation}
	[\zp, \zm] = -\frac{1}{2 \beta (H_*)^{-2} \epsilon_* (k_*)^{n_s - 1} c_s^{\beta - 2}} \delta^{3} (x - y) \ .
\end{equation}
Finally, we need to supplement the theory with initial conditions. The gaussian initial condition reads 
\begin{equation}
    \langle \zp \zp \rangle = \frac{H_*^4}{4 \Mpl^2 \dot H_* c_s} \frac{1}{k^3} \left(\frac{k_*}{k}\right)^{2 \epsilon + \eta + s} \ ,
\end{equation}
which is just the power spectrum.

\section{UV Calculations using the Mellin Representation}\label{app:mellin}

\subsection{Preliminaries}

We will be using the Hankel functions Mellin representation
\begin{equation}
	i \pi e^{i\pi \nu/2} H^{(1)}_{\nu} (z) = \int_{c-i\infty}^{c+i\infty}\frac{ds}{2\pi i} \Gamma \left ( s + \frac{\nu}{2} \right ) \Gamma \left ( s - \frac{\nu}{2} \right )\left ( -\frac{i z}{2} \right )^{-2s} \ ,
\end{equation}
for a suitable choice of $c$ (namely one that allows us to enclose all the poles of the $\Gamma$-functions). We apply this formula to the mode function for $\zeta$, 
\begin{equation}
\bar \zeta  = \Delta_\zeta \bar \zeta_c \qquad \bar \zeta_c = -i\sqrt{\frac{\pi}{2}} e^{\frac{1}{2}i \pi(\nu + 1/2)} \tau^{\nu} {\rm H}^{(1)}_\nu \left( - k \tau \right) \ ,
\end{equation}
with $2\nu= 3 + 2\epsilon + \kappa  = 3 - (n_s-1)$ and $c_s=1$. These mode functions appear throughout typical in-in calculations and are now expressed in Mellin space. In these variables, the problem of calculating loop corrections or relating the results back to the momentum and time, boils down to identification of poles in Mellin space. This representation of the Hankel function has left poles located at negative values of $s$. However, after performing loop integration, some poles will be introduced at positive values of $s$ as well. Nevertheless, as long as the poles are simple and non-overlapping, the regulated theory is one where we  choose a contour that keeps only the negative poles and we arrive at a convergent answer. As with other regulating schemes, this eliminates any power law divergences, but retains log divergences through the appearance of coinciding poles.

It will be important that our interactions of $\zeta$ always involve derivatives (as is required by symmetry). As a result, we are often not using $\zeta$ directly, but instead must take derivatives thereof. For spatial derivatives this is just multiplying by powers of $k\tau$. The behavior of time derivatives is less apparent, so we will derive it as follows. We start from the mode function
\begin{align}
	\bar \zeta_c &= -i e^{i(\nu + \frac{1}{2}) \frac{\pi}{2}} \frac{\sqrt{\pi}}{2} (-\tau)^{3/2 - (n_s - 1)/2}  H\left ( \frac{3}{2} - \frac{(n_s - 1)}{2}, k \tau \right )\\
	&= -\frac{1}{2\sqrt{\pi}} (-\tau)^{\frac{3 - (n_s - 1)}{2}} \int \frac{ds}{2\pi i} \Gamma \left ( s + \frac{3 - (n_s - 1)}{4} \right ) \Gamma \left ( s - \frac{3 - (n_s - 1)}{4} \right ) \left ( - \frac{ik \tau}{2} \right )^{-2s} \nonumber \ .
 \end{align}
A time derivative is just the logarithmic derivative with respect to $\tau$ and therefore we find
\begin{align}
	\dot{\bar \zeta}_c &= -\frac{1}{2\sqrt{\pi}} (-\tau)^{\frac{3 - (n_s - 1)}{2}} \int \frac{ds}{2\pi i} \left ( -2s + \frac{3}{2} - \frac{n_s - 1}{2} \right ) \cdot \Gamma  \left ( s + \frac{3 - (n_s - 1)}{4} \right ) \nonumber \\
	& \hspace{5cm} \Gamma \left ( s - \frac{3 - (n_s - 1)}{4} \right ) \left ( - \frac{ik \tau}{2} \right )^{-2s}\\
	&= \frac{1}{\sqrt{\pi}} (-\tau)^{\frac{3 - (n_s - 1)}{2}} \int \frac{ds}{2\pi i} \Gamma \left ( s + \frac{3 - (n_s - 1)}{4} \right ) \Gamma \left ( s + \frac{1 - (n_s - 1)}{4} \right ) \left ( - \frac{ik \tau}{2} \right )^{-2s} \nonumber \ .
\end{align}
It is important that the poles of both $\Gamma$ functions are located only at negative values of $s$. In the limit $k\tau \to 0$, this is just the statement that $\dot \zeta \to 0$, as we know from the EFT, but this is a statement that extends to any $k\tau$. This will be a key property of the Mellin representation that plays a very important role in the complete calculation.

We will compute the one loop power spectra
\begin{equation}
	\begin{aligned}
	\langle \zeta (\k) \zeta(\kp) \rangle &= i^2 \int d^4x \, [H_{\rm int}, [H_{\rm int}, \zeta(\k) \zeta(\kp)]]\\
	\end{aligned}
\end{equation}
where
\begin{equation}
 	S_{\rm int} = \int dt \, d^3x\, \bigg[ -\frac{2\lambda}{H^3} a^3 \dot{\zeta}_c^3 + \frac{\bar \lambda}{H^3} a \dot{\zeta}_c (\partial_i \zeta_c)^2 \bigg] \ .
\end{equation}
This expression is just the interaction picture expanded out to second order. Let us look at various permutations for this term, and analyze each of them separately via Mellin. Analytic continuation in the Mellin variables will achieve the same result as continuing to $t \to t(1\pm i \epsilon)$.

\subsection[One loop from Time Derivative Term]{One loop from $\dot{\zeta}^3$}

In the first example, we will include only the interaction
\begin{equation}
	H_{\rm int} = -{\cal L}_{\rm int} = - \lambda a^3 \dot{\zeta}^3 \ .
\end{equation}
In practice, calculating the commutators is less desirable that calculating the contributions from the different operator orderings. We will refer to the orderings in terms of how many factors of $H_{\rm int}$ appear on each side of the operator, so that the $(n,m)$-term has $n$ factors to the left and $m$ factors to the right.

\subsubsection{$(0,2)$ leg}

We will start with the term
\begin{equation}
	\langle \zeta(\k) \zeta(\kp) \rangle_{(0,2)} \equiv - \int_{-\infty}^{\tau} d\tau_1 \int_{-\infty}^{\tau_1} d \tau_2 \langle \zeta(\k) \zeta(\kp) H_{\rm int} (\tau_1) H_{\rm int} (\tau_2) \rangle ,
\end{equation}
which we will refer to as the $(0,2)$ piece. The other pieces are the $(1,1)$ and the $(2,0)$ pieces.

Now we can determine this contribution using Wick contractions and the mode functions to find
\begin{equation}
	\begin{aligned}
	\langle \zeta(\k) \zeta(\kp) \rangle &= -\lambda^2 \int_{-\infty}^{\tau} d\tau_1 a^4(\tau_1) \int_{-\infty}^{\tau_1} d\tau_2 a^4(\tau_2)\\
	 &\int \Pi_i \frac{d^3p_i}{(2\pi)^3} \delta^3(p_{123}) \delta^3(p_{456})\langle \zeta(\k) \zeta(\kp) \dot{\zeta}_{123} \dot{\zeta}_{456} \rangle
	\end{aligned}
\end{equation}
Here $\dot \zeta_{123}$ involves three factors of $\dot \zeta_c$ carrying momenta labeled by $1,2,3$. After the appropriate contractions, we get 
\begin{align}
 	\langle \zeta(\k) \zeta(\kp) \rangle &= -\lambda^2 \int_{-\infty}^{\tau} d\tau_1 a^4(\tau_1) \int_{-\infty}^{\tau_1} d\tau_2 a^4(\tau_2) \nonumber \\
	 &\int \frac{d^3p} {(2\pi)^3} \zeta(\k, \tau) \zeta(\kp, \tau) \dot{\zeta}_c^* (\k, \tau_1) \dot{\zeta}_c^* (\kp, \tau_2) \dot{\zeta}_c (\p, \tau_1) \dot{\zeta}_c^* (\p, \tau_2) \dot{\zeta}_c(\k - \p, \tau_1) \dot{\zeta}_c^* (\k - \p, \tau_2) \nonumber \\
	 &= - \lambda^2 \zeta(\k, \tau) \zeta(\kp, \tau) \int_{-\infty}^{\tau} d \tau_1 \, a^4(\tau_1) \int_{-\infty}^{\tau_1} d \tau_2 \, a^4(\tau_2) \int \, \frac{d^3p}{(2\pi)^3}  \\
  & \left( \prod_{i = 1}^{6} \int \frac{ds_i}{(2 \pi i)} \Gamma \left ( s_i + \frac{3 - (n_s - 1)}{4} \right ) \Gamma \left ( s_i + \frac{1 - (n_s - 1)}{4} \right )\left( - \frac{ip_i \tau_i}{2} \right)^{-2s_i} \right)\nonumber \\
	 & \qquad \times \frac{1}{\pi^{3}}(-\tau_1)^{\frac{9 - 3(n_s - 1)}{2}} (-\tau_2)^{\frac{9 - 3(n_s - 1)}{2}}  (-1)^{-2s_1 - 2s_2 - 2s_4  -2s_6} \ .\nonumber
	\end{align}	
 We want to simplify the integrals now. The advantage of Mellin comes from the fact that the $p$ and $\tau$ integrals are now separately scaleless, which allows us to deal with them effectively in a dim reg manner. Doing the momentum integral, we get,
\begin{equation}
	\int \frac{d^3p}{(2\pi)^3} p^{-2(s_3 + s_4)} (k - p)^{-2(s_5 + s_6)}
\end{equation}
We take the limit where $p \gg k$, or where the loop momenta is much greater than the internal momenta. In the other region, $p \ll k$, one can use the consistency conditions to prove that the integral only gives slow roll suppressed contributions. Hence, we have
\begin{equation}
	\int \frac{d^3p}{(2\pi)^3} p^{-2(s_3 + s_4 + s_5 + s_6)} \to -\frac{i}{2\pi} \delta \left ( \frac{3}{2} - (s_3 + s_4 + s_5 + s_6) \right ) \ ,
\end{equation}
where the $\delta$-function is understood as arising from a pole in the Mellin variables that we must include when performing future integrals. The result of including this pole shifts some of the poles from our $\Gamma$-functions outside of  the contour of integration. 

Having dealt with the momenta integral, we now turn to the time integral. We first integrate over $\tau_2$ to get
\begin{equation}
\begin{aligned}
	\int_{-\infty}^{\tau_1} &d \tau_2 \,\, a^4(\tau_2) \, (\tau_2)^{\frac{9 - 3(n_s - 1)}{2}} (\tau_2)^{-2(s_2 + s_4 + s_6)}\\
	 &= \frac{2 (\tau_1)^{\frac{3 - 3(n_s - 1)}{2} - 2(s_2 + s_4 + s_6)}}{3 - 3(n_s - 1) - 4(s_2 + s_4 + s_6)}  \ ,
\end{aligned}
\end{equation}
and then perform the $\tau_1$ integral
\begin{equation}
\begin{aligned}
&\int_{-\infty}^{\tau} d \tau_1 \frac{2(\tau_1)^{\frac{1 - 3(n_s - 1)}{2} - 2(s_1 + s_3 + s_5)} (\tau_1)^{\frac{3 - 3(n_s - 1)}{2} - 2(s_2 + s_4 + s_6)}}{3 - 3(n_s - 1) - 4(s_2 + s_4 + s_6)}\\
	&= \frac{2\tau^{3 - 3(n_s - 1) - 2(s_1 + s_2 + \hdots + s_6)}}{(3 - 3(n_s - 1) - 2(s_1 + s_2 + \hdots + s_6))(3 - 3(n_s - 1) - 4(s_2 + s_4 + s_6))}
\end{aligned}
\end{equation}
As $\tau \to 0$, this turns out to be a $\delta$ function as well,
\begin{equation}\label{eq:tau0limit}
	\frac{\tau^{3 - 3(n_s - 1) - 2(s_1 + s_2 + \hdots + s_6)}}{3 - 3(n_s - 1) - 2(s_1 + s_2 + \hdots + s_6)} \to i \pi \delta \left ( \frac{3}{2} - (s_1 + s_2 + \hdots + s_6)  \right ) \ ,
\end{equation}
enforcing the fact that we are only picking out contributions that survive in the late future. All together, this means the poles in Mellin space after integration are determined by the expression
\begin{equation}
\begin{aligned}
	&\frac{1}{3 - 3(n_s - 1) - 4(s_2 + s_4 + s_6)} \delta \left ( \frac{3}{2} - (s_1 + s_2 + \hdots + s_6)  \right ) \\
	&\prod_{i} \Gamma \left ( s_i + \frac{3 - (n_s - 1)}{4} \right ) \Gamma \left ( s_i + \frac{1 - (n_s - 1)}{4} \right ) \delta \left ( \frac{3}{2} - s_3 - s_4 - s_5 - s_6 \right )\\
&\to \frac{1}{3 - 3(n_s - 1) - 4(s_2 + s_4 + s_6)} \delta (s_1 + s_2) \\
&\prod_{i}\Gamma \left ( s_i + \frac{3 - (n_s - 1)}{4} \right ) \Gamma \left ( s_i + \frac{1 - (n_s - 1)}{4} \right ) \delta \left ( \frac{3}{2} - s_3 - s_4 - s_5 - s_6 \right ) \ .
\end{aligned}
\end{equation}
Notice that in the last step, because of the loop momentum, a non-vanishing results as $\tau \to 0$ requires $s_1 +s_2 =0$. We will focus on the poles of this expression, rather than the full 1-loop result, as this is what is relevant for determining the divergences in the final result.

In Mellin space, divergences in these loop calculations manifest themselves as collisions between two or more poles. A properly regulated theory should only have simple poles and the collision of the poles when the regulator is removed gives rise to logarithmic divergences (similar to dimensional regularization). We will show that this doesn't arise, as expected from the EFT description. To see this, we can first integrate over $s_1$ using the $\delta$ function to set $s_1 = -s_2$. We can also integrate out $s_5$ and set $s_5 = 3/2 - s_3 - s_4 - s_6$. Then we have
\begin{equation}
\begin{aligned}
	&\frac{1}{3 - 3(n_s - 1) - 4(s_2 + s_4 + s_6)} \Gamma \left ( s_2 + \frac{3 - (n_s - 1)}{4} \right ) \Gamma \left ( s_2 + \frac{1 - (n_s - 1)}{4} \right )\\
	&\cdot \Gamma \left ( -s_2 + \frac{3 - (n_s - 1)}{4} \right ) \Gamma \left ( -s_2 + \frac{1 - (n_s - 1)}{4} \right ) \cdot \Gamma_{s_3, s_4, s_6} \\
	&\cdot \Gamma \left (-s_3 - s_4 - s_6 + \frac{9 - (n_s - 1)}{4} \right ) \Gamma \left (-s_3 - s_4 - s_6 + \frac{7 - (n_s - 1)}{4} \right )  \ .
\end{aligned}
\end{equation}
Now, we have some poles on the negative part of the number line and some poles on the positive part of the number line due to the $\Gamma$ function. It is clear from inspection that the left and right poles are well separated, and a line going through $s^{\cal C}_i = 0$ is all that is required. The key point was that due to the time derivative, $\dot{\zeta}^3$ has no poles on the positive number line. This ensures that none of the $s_2, s_3, s_4, s_6$ poles can gives rise to double poles in the expression above.

We now want to calculate the finite part of the sum. We take $n_s - 1 = 0$ to simplify the calculation. The pole structure we care about is
\begin{equation}
	\begin{aligned}
	&\frac{-(1)^{-2s_4 - 2s_6}}{3 - 3(n_s - 1) - 4(s_2 + s_4 + s_6)} \Gamma \left ( s_2 + \frac{3 - (n_s - 1)}{4} \right ) \Gamma \left ( s_2 + \frac{1 - (n_s - 1)}{4} \right )\\
	&\cdot \Gamma \left ( -s_2 + \frac{3 - (n_s - 1)}{4} \right ) \Gamma \left ( -s_2 + \frac{1 - (n_s - 1)}{4} \right ) \cdot \Gamma_{s_3, s_4, s_6} \\
	&\cdot \Gamma \left (-s_3 - s_4 - s_6 + \frac{9 - (n_s - 1)}{4} \right ) \Gamma \left (-s_3 - s_4 - s_6 + \frac{7 - (n_s - 1)}{4} \right ) \ .
\end{aligned}
\end{equation}
The first thing we sum is the $s_3$ poles, since it only has $\Gamma$ matrix structure,
\begin{equation}
	\begin{aligned}
	\int \frac{ds_3}{(2\pi i)}&\Gamma \left (s_3 + \frac{3}{4} \right ) \Gamma \left (s_3 + \frac{1}{4} \right ) \Gamma \left (-s_3 - s_4 - s_6 + \frac{9}{4} \right ) \Gamma \left (-s_3 - s_4 - s_6 + \frac{7 }{4} \right ) \\
	&= 2^{-7 + 4(s_4 + s_6)} \pi \Gamma(4 - 2(s_4 + s_6)) \ .
	\end{aligned}
\end{equation}
Let us turn to $s_2$ poles next, which give us,
\begin{equation}
	\begin{aligned}
	&\frac{1}{3 - 4(s_2 + s_4 + s_6)}\Gamma \left ( s_2 + \frac{3}{4} \right ) \Gamma \left ( s_2 + \frac{1}{4} \right ) \Gamma \left ( -s_2 + \frac{3}{4} \right ) \Gamma \left ( -s_2 + \frac{1}{4} \right )\\
	&= 2\pi \frac{_p F_q \bigg[\left (1, \frac{3}{2} - (s_4 + s_6), 1 - (s_4 + s_6)  \right ), \left ( 2 - (s_4 + s_6), \frac{5}{2} - (s_4 + s_6) \right ), 1 \bigg]}{(4 - 4(s_4 + s_6))(6 - 4(s_4 + s_6))}\\
	&= \frac{\pi}{4}\bigg[ (H_n\left ( \frac{3}{4} - (s_4 + s_6) \right ) - H_n(-(s_4 + s_6) ) \bigg] \ , 
 	\end{aligned} 	
\end{equation} 
where $H_n$ is the Harmonic numbers, defined as
\begin{equation}
	H_n(x) = \sum_{i = 0}^{x} \frac{1}{i} \ .
\end{equation}
Putting these together, we get that the loop integral gives us
\begin{equation}
\begin{aligned}
	\frac{\langle \zeta(\k) \zeta(\kp)\rangle}{P_\zeta} =   -\frac{\lambda^2}{ 2^{10} \pi^3}& \int \frac{ds_4 ds_6}{(2\pi i )^2} (-1)^{-2s_4 - 2s_6} 2^{4(s_4 + s_6)} \Gamma(4 - 2(s_4 + s_6)) \\
    &\bigg[ H_n\left ( \frac{3}{4} - (s_4 + s_6) \right ) - H_n(-(s_4 + s_6) \bigg] \\
    &\Gamma \left (s_4 + \frac{3}{4} \right ) \Gamma \left (s_4 + \frac{1}{4} \right ) \Gamma \left (s_6 + \frac{3}{4} \right ) \Gamma \left (s_6 + \frac{1}{4} \right ) \ .
\end{aligned}
\end{equation}
While we are unable to find a analytic sum over the poles of $s_4, s_6$, we can perform a numerical sum. Concretely, we can define 
\begin{equation}
\begin{aligned}
    \xi = \int \frac{ds_4 ds_6}{(2\pi i )^2} &(-1)^{-2s_4 - 2s_6} 2^{4(s_4 + s_6)} \bigg[ (H_n\left ( \frac{3}{4} - (s_4 + s_6) \right ) - H_n(-(s_4 + s_6) ) \bigg] \\
    &\Gamma(4 - 2(s_4 + s_6))\Gamma \left (s_4 + \frac{3}{4} \right ) \Gamma \left (s_4 + \frac{1}{4} \right ) \Gamma \left (s_6 + \frac{3}{4} \right ) \Gamma \left (s_6 + \frac{1}{4} \right ) \ .
\end{aligned}
\end{equation}
First, we compute the first $n_1$ poles of $s_4$ coming from $\Gamma(s_4 + 3/4)$. We call this function $\xi_1(n_1)$. The sum of reside of these poles give us
\begin{equation}
    \begin{aligned}
        \xi_1(n_1) &= \sum_{m = 1}^{n_1}\int \frac{ds_6}{2\pi i}  (-1)^{-2s_6} 2^{4 s_6} \Gamma \left (s_6 + \frac{3}{4} \right ) \Gamma \left (s_6 + \frac{1}{4} \right )\\
        &\bigg[ (-1)^{3/2 + 2m}\frac{(-1)^m}{m!} 2^{-3 - 4m} \bigg( H_n\left ( \frac{6}{4} + m - s_6 \right ) - H_n\left(\frac{3}{4} + m - s_6 \right) \bigg) \\
        &\cdot \Gamma \left(-m - \frac{1}{2} \right)\Gamma \left( \frac{11}{2} + 2m - 2s_6 \right) \bigg]\\
        &= \int \frac{ds_6}{2\pi i}  (-1)^{-2s_6} 2^{4 s_6} \Gamma \left (s_6 + \frac{3}{4} \right ) \Gamma \left (s_6 + \frac{1}{4} \right ) \sum_{m=1}^{n_1}f_1(m) \ .
    \end{aligned}
\end{equation}
Now, we sum the first $n_1$ poles of $s_4$ coming from $\Gamma(s_4 + 1/4)$. We call this function $\xi_2(n_1)$. We get that 
\begin{equation}
    \begin{aligned}
        \xi_2(n_1) &= \sum_{m = 1}^{n_1}\int \frac{ds_6}{2\pi i}  (-1)^{-2s_6} 2^{4 s_6} \Gamma \left (s_6 + \frac{3}{4} \right ) \Gamma \left (s_6 + \frac{1}{4} \right )\\
        &\bigg[ (-1)^{1/2 + 2m}\frac{(-1)^m}{m!} 2^{-1 - 4m} \bigg( H_n\left ( 1 + m - s_6 \right ) - H_n\left(\frac{1}{4} +  m - s_6 \right) \bigg) \\
        &\cdot \Gamma \left(-m + \frac{1}{2} \right)\Gamma \left( \frac{9}{2} + 2m - 2s_6 \right) \bigg]\\
        &= \int \frac{ds_6}{2\pi i}  (-1)^{-2s_6} 2^{4 s_6} \Gamma \left (s_6 + \frac{3}{4} \right ) \Gamma \left (s_6 + \frac{1}{4} \right ) \sum_{m=1}^{n_1}f_2(m) \ .
    \end{aligned}
\end{equation}
We of course have that $\xi = \lim_{n_1 \to \infty} \big[\xi_1(n_1) + \xi_2(n_2)\big]$. Now, we want to simplify $\xi_1$ and $\xi_2$ further, by also summing over the $s_6$ poles. Because $s_4, s_6$ were symmetric, let us also sum over the first $n_1$ poles for $s_6$. We first sum over the $\Gamma(s_6 + 3/4)$ poles, and then  $\Gamma(s_6 + 1/4)$ poles. Let us define the new functions as 
\begin{equation}
\begin{aligned}
    \xi'_1(n_1) &= \sum_{l = 1}^{n_1}  (-1)^{3/2 + 2l} 2^{-3 - 4l} \frac{(-1)^{l}}{l!} \Gamma \left (-l - \frac{1}{2} \right ) \left( \sum_{m=1}^{n_1}f_1(m) + f_2(m) \right)\\
    \xi'_2(n_1) &= \sum_{l = 1}^{n_1}  (-1)^{1/2 + 2l} 2^{-1 - 4l} \frac{(-1)^{l}}{l!} \Gamma \left (-l + \frac{1}{2} \right ) \left( \sum_{m=1}^{n_1}f_1(m) + f_2(m) \right) \ .
\end{aligned}
\end{equation}
Finally, then, we have that
\begin{equation}
    \xi = \lim_{n \to \infty} (\xi_1'(n) + \xi_2'(n)) \ .
\end{equation}
Now, this form makes it conducive to numerically evaluate it. We comment that we see the sum start to asymptote at around $n_1 = 120$, to a value of 
\begin{equation} 
    \xi \approx -2.972 \times 10^9 \ .
\end{equation}
Then, the loop result is given by
\begin{equation}
 \langle \zeta(\k) \zeta(\kp) \rangle_{\rm 1-loop}'=   P_\zeta(k) \bigg[-\frac{\lambda^2}{2^{10}\pi^3} \xi\bigg] = -9.36 \times 10^4 \lambda^2 P_\zeta \ .
\end{equation}
We just comment that this is a scheme dependent finite contribution. It is important to point out that this is not equivalent to the result of~\cite{Kristiano:2022zpn}, which point $\xi \propto 1/(n_s-1)$ which would be divergent in this case. While we would still like to have a better understanding of why this constant is large, it does not appear to signal a deeper issue with perturbation theory. 

As we have seen, we chose contours such that $s_1 = -s_2$, and we got a finite scheme dependent result. This counterterm can be absorbed into the 6 derivative operators. Choosing contours, however, is equivalent to choosing schemes. We could have just recognized, when we encountered $(k \tau)^{-2s_1 - 2s_2}$, that since all the $\Gamma$ function poles lie on the negative axis, we can never get a function that behaves like $\tau^0$. Thus, if we do not enforce the $s_1 = -s_2$ constraint coming from the time integral (which we are not forced to do anyways), we will immediately see that no contribution survives in the $\tau \to 0$ limit, and we get that the loop integral is 0, as we would expect from a nice dim reg calculation to regulate power law divergences. Regardless of whether we chose a scheme where we get a finite contribution or 0, we see that no $\log$ will be generated, and hence Mellin shows that the loop is regulated without any need for RG.

\subsubsection{$(2,0)$ leg}

The $(2,0)$ will be given by the complex conjugate of the above calculation. Since we have already established that the above calculation has no poles, the $(2,0)$ legs will have no poles either. It will again give the same finite scheme dependent answer from the loop.

\subsubsection{$(1,1)$ leg}

Let us calculate the $(1,1)$ leg. This will be given by
\begin{equation}
\begin{aligned}
	\langle \zeta(\k) \zeta(\kp) \rangle' &= - \int_{-\infty}^{\tau} d\tau_1 \int_{-\infty}^{\tau} d \tau_2 \langle H_{\rm int} (\tau_1) \zeta(\k) \zeta(\kp) H_{\rm int} (\tau_2) \rangle' .\\
	 &= -\lambda^2 \int_{-\infty}^{\tau} d\tau_1 a^4(\tau_1) \int_{-\infty}^{\tau} d\tau_2 a^4(\tau_2)\\
	 &\int \Pi_i \frac{d^3p_i}{(2\pi)^3} \delta^3(p_{123}) \delta^3(p_{456})\langle \dot{\zeta}_{123} \zeta(\k) \zeta(\kp) \dot{\zeta}_{456} \rangle
\end{aligned}
\end{equation}
After Wick contractions, we will get 
	\begin{align}
 	\langle \zeta(\k) \zeta(\kp) \rangle &= -\lambda^2 \int_{-\infty}^{\tau} d\tau_1 a^4(\tau_1) \int_{-\infty}^{\tau} d\tau_2 a^4(\tau_2) \nonumber \\
	 &\int \frac{d^3p} {(2\pi)^3} \dot{\zeta}_c (\k, \tau_1) \zeta^* (\k, \tau) \zeta(\k, \tau) \dot{\zeta}_c^* (\k, \tau_2) \dot{\zeta}_c (\p, \tau_1) \dot{\zeta}_c^* (\p, \tau_2) \dot{\zeta}_c(\k - \p, \tau_1) \dot{\zeta}_c^* (\k - \p, \tau_2)  \nonumber\\
	 &= - \lambda^2 \zeta(\k, \tau) \zeta^*(\kp, \tau) \int_{-\infty}^{\tau} d \tau_1 \, a^4(\tau_1) \int_{-\infty}^{\tau} d \tau_2 \, a^4(\tau_2) \int \, \frac{d^3p}{(2\pi)^3}\\
  & \prod_{i = 1}^{6} \int \frac{ds_i}{(2 \pi i)} \Gamma \left ( s_i + \frac{3 - (n_s - 1)}{4} \right ) \Gamma \left ( s_i + \frac{1 - (n_s - 1)}{4} \right ) \left ( - \frac{i p_i \tau_i}{2} \right )^{-2s_i} \nonumber \\
	 & (-1)^{- 2s_2 - 2s_4  -2s_6}  \frac{1}{\pi^{3}}(-\tau_1)^{\frac{9 - 3(n_s - 1)}{2}} (-\tau_2)^{\frac{9 - 3(n_s - 1)}{2}}  \nonumber
	\end{align}	
Compared to the $(2,0)$ leg, the major difference is in the time integral. Instead of a nested time integral, we get two time integrals running from $- \infty \to \tau$. Thus we have
\begin{equation}
\begin{aligned}
    \int_{-\infty}^{\tau} d \tau_2 \,\, a^4(\tau_2) \, (\tau_2)^{\frac{9 - 3(n_s - 1)}{2}} (\tau_2)^{-2(s_2 + s_4 + s_6)} =  \frac{2 (\tau)^{\frac{3 - 3(n_s - 1)}{2} - 2(s_2 + s_4 + s_6)}}{3 - 3(n_s - 1) - 4(s_2 + s_4 + s_6)}\ ,
\end{aligned}
\end{equation}
and the same for the $\tau_1$ integral. The momenta integral remains the same,
\begin{equation}
	\int \frac{d^3p}{(2\pi)^3} p^{-2(s_3 + s_4 + s_5 + s_6)} \to -\frac{i}{2\pi} \delta \left ( \frac{3}{2} - (s_3 + s_4 + s_5 + s_6) \right ) \ .
\end{equation}
Now, we see that since we have all the poles on the negative axis, we again cannot get any time dependence in the $\tau \to 0$ limit. Mellin again sets the this term to 0.

\subsection[One loop from Gradient Term]{One loop from $\dot{\zeta} (\partial \zeta)^2$}

Now we would like to repeat the above analysis of one-loop power spectrum for the $\bar \lambda a \dot{\zeta}_c (\partial_i \zeta_c)^2$ interaction. We will focus only on showing the absence of divergences. We focus on the $(2,0)$ 1-loop term, written schematically as
\begin{equation}
	\begin{aligned}
	\langle \zeta(\k) \zeta(\kp) \rangle &= -\bar\lambda^2 \int_{-\infty}^{\tau} d\tau_1 a^2(\tau_1) \int_{-\infty}^{\tau_1} d\tau_2 a^2(\tau_2)\\
	 &\int \Pi_i \frac{d^3p_i}{(2\pi)^3} \delta^3(k_{123}) \delta^3(k_{456})\langle \zeta(p) \zeta(p') \dot{\zeta}_{1} (\partial_i \zeta)^2_{23} \dot{\zeta}_{4} (\partial_i \zeta)^2_{56} \rangle \ .
	\end{aligned}
\end{equation}
As we saw in the previous example, the other $(m,n)$ terms have the same pole structure, and hence showing that poles are well separated in this example will take care of all the poles. Of course, we have multiple Wick contractions that are possible. We explain in detail the pole structure of two possible Wick contractions.

\subsubsection{First Wick contraction}

The first contraction will focus on terms where both the $\dot{\zeta}$ contracts with the external $\zeta(\k), \zeta(k')$. Then, we get
\begin{align}
 	\langle \zeta(\k) \zeta(\kp) \rangle &= -\bar \lambda^2 \int_{-\infty}^{\tau} d\tau_1 a^2(\tau_1) \int_{-\infty}^{\tau_1} d\tau_2 a^2(\tau_2) \, (k \cdot ( k - p) )^2  \nonumber \\
	 &\int \frac{d^3p} {(2\pi)^3} \zeta(\k, \tau) \zeta(\kp, \tau) \dot{\zeta}_c^* (\k, \tau_1) \dot{\zeta}_c^* (\kp, \tau_2) \zeta_c(\p, \tau_1) \zeta_c^* (\p, \tau_2) \zeta_c(\k - \p, \tau_1) \zeta_c^* (\k - \p, \tau_2)  \nonumber\\
	 &= - \bar \lambda^2 \zeta(\k, \tau) \zeta(\kp, \tau) \int_{-\infty}^{\tau} d \tau_1 \, a^2(\tau_1) \int_{-\infty}^{\tau_1} d \tau_2 \, a^2(\tau_2) \int \, \frac{d^3p}{(2\pi)^3} \, \nonumber \\
	 &\left(\prod_{i = 1}^{6} \int \frac{ds_i}{(2 \pi i)}\Gamma \left ( s_i + \frac{3 - (n_s - 1)}{4} \right ) \Gamma \left ( s_i - \frac{3 - (n_s - 1)}{4} \right ) \left ( - \frac{ip_i \tau_i}{2} \right )^{-2s_i} \right)  \nonumber \\
	  &(-1)^{-2s_1 - 2s_2 - 2s_4 - 2s_6} (-\tau_1)^{\frac{9 - 3(n_s - 1)}{2}} (-\tau_2)^{\frac{9 - 3(n_s - 1)}{2}} (k \cdot ( k - p) )^2 \ .
\end{align}
As before, we can evaluate the loop integral assuming the internal momentum is hard ($p \gg k)$, so that
\begin{equation}
	\int \frac{d^3p}{(2\pi)^3} p^{4-2(s_3 + s_4 + s_5 + s_6)} \to -\frac{i}{2\pi} \delta \left ( \frac{7}{2} - s_3 + s_4 + s_5 + s_6 \right ) \ .
\end{equation}
We also calculate the integrals as before, to find
\begin{equation}
\begin{aligned}
	\int_{-\infty}^{\tau_1} &d \tau_2 \,\, a^2(\tau_2) \, (\tau_2)^{\frac{9 - 3(n_s - 1)}{2}} (\tau_2)^{-2(s_2 + s_4 + s_6)}\\
	 &= \frac{2 (\tau_1)^{\frac{7 - 3(n_s - 1)}{2} - 2(s_2 + s_4 + s_6)}}{7 - 3(n_s - 1) - 4(s_2 + s_4 + s_6)} \ ,
\end{aligned}
\end{equation}
and
\begin{equation}
\begin{aligned}
	\int_{-\infty}^{\tau} d \tau_1 \, &a^2(\tau_1) (\tau_1)^{\frac{9 - 3(n_s - 1)}{2}} (\tau_1)^{-2(s_1 + s_3 + s_5)}\\
	&= \frac{\tau^{7 - 3(n_s - 1) - 2(s_1 + s_2 + \hdots + s_6)}}{7 - 3(n_s - 1) - 2(s_1 + s_2 + \hdots + s_6)} \ .
\end{aligned}
\end{equation}
Using the $\delta$ function from the momenta integral, we again reduce to a $\tau^{-2(s_1 + s_2)}$ dependence. Since $s_1, s_2$ are the poles associated with the $\dot{\zeta}$ term, these poles are negative and the integral is set to 0.

\subsubsection{Other Wick contraction}

To convince the reader that the same pattern holds, let us look at the term where the $\partial_i \zeta$ contracts with the external $\zeta$. We get 
\begin{align}
 	\langle \zeta(\k) \zeta(\kp) \rangle &= -
  \bar \lambda^2 \int_{-\infty}^{\tau} d\tau_1 a^2(\tau_1) \int_{-\infty}^{\tau_1} d\tau_2 a^2(\tau_2) \, (p \cdot k)^2 \nonumber \\
	 &\int \frac{d^3k} {(2\pi)^3} \zeta(\k, \tau) \zeta(\kp, \tau) \dot{\zeta}_c(\k - \p, \tau_1) \dot{\zeta}^*_c (\k - \p, \tau_2)  \zeta^*_c  (\p, \tau_1) \zeta^*_c (\p, \tau_2) \zeta_c(\k, \tau_1) \zeta^*_c (\k, \tau_2) \nonumber \\
	 &= - \bar \lambda^2 \zeta(\p, \tau) \zeta(\p, \tau) \int_{-\infty}^{\tau} d \tau_1 \, a^2(\tau_1) \int_{-\infty}^{\tau_1} d \tau_2 \, a^2(\tau_2) \, \nonumber \\
	 & \int \, \frac{d^3k}{(2\pi)^3} (-\tau_1)^{\frac{9 - 3(n_s - 1)}{2}} (-\tau_2)^{\frac{9 - 3(n_s - 1)}{2}} (p \cdot k)^2\\
	  & \prod_{i = 1}^{6} (-1)^{-2s_2 - 2s_3 - 2s_4 - 2s_6} \int \frac{ds_i}{(2 \pi i)} \Gamma \left ( s_i + \frac{3 - (n_s - 1)}{4} \right )\nonumber \\ 
	  & \qquad \Gamma \left ( s_i - \frac{3 - (n_s - 1)}{4} \bigg|  s_i + \frac{1 - (n_s - 1)}{4} \right ) \left ( - \frac{ik_i \tau_i}{2} \right )^{-2s_i}\nonumber \ ,
\end{align}	
where $\Gamma(a | b)$ picks the appropriate $\Gamma$ function $\Gamma(a)$ or $\Gamma(b)$ for the given $s_i$, depending on whether it is a $\dot{\zeta}$ or $\zeta$. Now, it is time to do the integrals in various fashion. We will first start with the momenta integral, which reads
\begin{equation}
	\int \frac{d^3p}{(2\pi)^3} p^{2-2(s_3 + s_4)} (k - p)^{-2(s_5 + s_6)} \ .
\end{equation}
We take the hard limit, $(p \gg k)$, giving us
\begin{equation}
	\int \frac{d^3p}{(2\pi)^3}p^{2-2(s_3 + s_4 + s_5 + s_6)} \to -\frac{i}{2\pi} \delta \left ( \frac{5}{2} - (s_3 + s_4 + s_5 + s_6) \right ) \ .
\end{equation}
Now, we proceed to do the time integral. The $\tau_2$ integral gives us
\begin{equation}
\begin{aligned}
	\int_{-\infty}^{\tau_1} &d \tau_2 \,\, a^2(\tau_2) \, (\tau_2)^{\frac{9 - 3(n_s - 1)}{2}} (\tau_2)^{-2(s_2 + s_4 + s_6)}\\
	 &= \frac{2 (\tau_1)^{\frac{7 - 3(n_s - 1)}{2} - 2(s_2 + s_4 + s_6)}}{7 - 3(n_s - 1) - 4(s_2 + s_4 + s_6)} 
\end{aligned}
\end{equation}
as from before. The $\tau_1$ integral then gives us
\begin{equation}
\begin{aligned}
	\int_{-\infty}^{\tau} d \tau_1 \, &a^2(\tau_1) (\tau_1)^{\frac{9 - 3(n_s - 1)}{2}} (\tau_1)^{-2(s_1 + s_3 + s_5)}\\
	&= \frac{\tau^{7 - 3(n_s - 1) - 2(s_1 + s_2 + \hdots + s_6)}}{7 - 3(n_s - 1) - 2(s_1 + s_2 + \hdots + s_6)} \ .
\end{aligned}
\end{equation}
Again, we have that enforcing the momentum integral $\delta$ function, we get $(k \tau)^{1 - 2(s_1 + s_2)}$ which is set to 0 by our Mellin prescription in the $\tau \to 0$ limit.

\section{Flat Spacetime Example} \label{app:flat}
Let us take a flat spacetime example to understand the renormalization to the propagator due to a  $\dot{\phi}^3$ interaction term. Modes deep inside the de Sitter horizon effectively behave as if they are in flat spacetime, so understanding the field in flat spacetime will guide our de Sitter result as well. The Lagrangian we consider is 
\begin{equation}
    {\cal L} = \dot{\phi}^2 - (\partial_i \phi)^2 + \frac{g}{\Lambda^2}\dot{\phi}^3 \ ,
\end{equation}
where $\Lambda$ is UV scale where the EFT description breaks down. Now, we calculate the corrections to the in-in propagator. The interaction picture operator in flat space is given by
\begin{equation}
    \phi(\k) = \int \frac{d^3k}{(2\pi)^3} \frac{1}{\sqrt{k}} \left( a_\k e^{ikx - ik t} + a_\k^\dagger e^{-ikx + ikt} \right) \ .
\end{equation}
At lowest order, the correction to $\langle \phi(\k) \phi(\kp) \rangle$ is given by
\begin{equation}
	\langle \phi(\k) \phi(\kp) \rangle = i^2 \int d^3x_1\, dt_1\, \int d^3x_2\, dt_2\, \big[H_{\rm int}(x_2, t), [H_{\rm int}(x_1, t), \phi(\k) \phi(\kp)] \big]  \ ,
\end{equation}
where $H_{\rm int} = \frac{g}{\Lambda^2}\dot{\phi}^3$. Considering the $(0,2)$ contraction, we have
\begin{equation}
\begin{aligned}
	\langle \phi(\k) \phi(\kp) \rangle &= - \int_{-\infty}^{t} dt_1 \int_{-\infty}^{t_1} d t_2 \langle \phi(\k) \phi(\kp) H_{\rm int} (t_1) H_{\rm int} (t_2) \rangle \\
	&= -\frac{g^2}{\Lambda^4} e^{-2ikt} \int \frac{d^3p}{(2\pi)^3} \frac{p^2 q^2 k^2}{k^2 q p} \int_{-\infty}^{t} dt_1 e^{i(k - q - p)t_1}\int_{-\infty}^{t_1} dt_2 e^{i(k + q + p)t_2} \ ,
\end{aligned} 
\end{equation}
where $q = |\p - \k|$. Doing the time integrals, we get
\begin{equation}
	\langle \phi(\k) \phi(\kp) \rangle = \frac{g^2}{\Lambda^4} \int \frac{d^3p}{(2\pi)^3} \frac{qk}{2k(k + q + p)}  \ .
\end{equation}
The free theory correlation function is given by
\begin{equation}
	\langle \phi(k) \phi(k') \rangle = \frac{1}{k} \ .
\end{equation}
Now, just from dimensional analysis, since the only scale in the $p$ integral is $k$, we will have that, up to scaleless integrals, the term that will be produced, when expanded in $p/k$ is (large loop momenta)
\begin{equation}
	\langle \phi(\k) \phi(\kp) \rangle \sim  \left (\alpha \frac{g^2 k^3}{\Lambda^4} \int \frac{d^3p}{(2\pi)^3} \frac{1}{p^3}  \right ) \ ,
\end{equation}
where $\alpha$ is some number. Similarly, if we look at the $(1,1)$ leg, we will get that the loop correction is given by 
\begin{equation}
\begin{aligned}
    \langle \phi(\k) \phi(\kp) \rangle &= - \int_{-\infty}^{t} dt_1 \int_{-\infty}^{t_1} d t_2 \langle \phi(\k) \phi(\kp) H_{\rm int} (t_1) H_{\rm int} (t_2) \rangle\\
    &= \frac{g^2}{\Lambda^4} \int \frac{d^3p}{(2\pi)^3} \frac{qk}{(k + q + p)^2} \ .
\end{aligned}
\end{equation}
Similarly, for large internal momenta ($p \gg k$), we can extract a term
\begin{equation}
    \langle \phi(\k) \phi(\kp) \rangle \sim  \left (\beta \frac{g^2 k^3}{\Lambda^4} \int \frac{d^3p}{(2\pi)^3} \frac{1}{p^3}  \right ) \ .
\end{equation}
We can guess the local operator that absorbs these counterterms. Let us take the interaction term
\begin{equation}
	S_{\rm int} = \int d^4x\, \frac{g^2}{\Lambda^4} \phi \partial_t^6 \phi \ . 
\end{equation}
This produces a correction of the form
\begin{equation}
	\langle \phi(\k) \phi(\kp) \rangle = i\frac{g^2}{\Lambda^4} \frac{e^{-2ikt}}{k^2} \cdot k^6 \cdot \int dt\, e^{2i kt} = \frac{g^2}{\Lambda^4} k^3 \ ,
\end{equation}
which is exactly the structure required to absorb the divergence from the loop integral. Note that similarly, we can also have $\phi \partial_i^6 \phi$ as a counterterm (these are the same operators, as one can see using the EOM). This shows that the loops generated by our interaction Lagrangian are absorbed by 6 derivative counterterms.

\clearpage
\phantomsection
\addcontentsline{toc}{section}{References}
\small
\bibliographystyle{utphys}
\bibliography{Refs}

\providecommand{\href}[2]{#2}\begingroup\raggedright\begin{thebibliography}{100}

\bibitem{Baumann:2009ds}
D.~Baumann,
  \href{http://dx.doi.org/10.1142/9789814327183_0010}{``{Inflation},''} in {\em
  {Theoretical Advanced Study Institute in Elementary Particle Physics}:
  {Physics of the Large and the Small}}, pp.~523--686.
\newblock 2011.
\newblock \href{http://arxiv.org/abs/0907.5424}{{\ttfamily arXiv:0907.5424
  [hep-th]}}.

\bibitem{Achucarro:2022qrl}
A.~Ach\'ucarro {\em et~al.}, ``{Inflation: Theory and Observations},''
  \href{http://arxiv.org/abs/2203.08128}{{\ttfamily arXiv:2203.08128
  [astro-ph.CO]}}.

\bibitem{Green:2022bre}
D.~Green, ``{Cosmic Signals of Fundamental Physics},''
  \href{http://dx.doi.org/10.22323/1.439.0005}{{\em PoS} {\bfseries TASI2022}
  (2024) 005}, \href{http://arxiv.org/abs/2212.08685}{{\ttfamily
  arXiv:2212.08685 [hep-ph]}}.

\bibitem{SimonsObservatory:2018koc}
{\bfseries Simons Observatory} Collaboration, P.~Ade {\em et~al.}, ``{The
  Simons Observatory: Science goals and forecasts},''
  \href{http://dx.doi.org/10.1088/1475-7516/2019/02/056}{{\em JCAP} {\bfseries
  02} (2019) 056}, \href{http://arxiv.org/abs/1808.07445}{{\ttfamily
  arXiv:1808.07445 [astro-ph.CO]}}.

\bibitem{CMB-S4:2020lpa}
{\bfseries CMB-S4} Collaboration, K.~Abazajian {\em et~al.}, ``{CMB-S4:
  Forecasting Constraints on Primordial Gravitational Waves},''
  \href{http://dx.doi.org/10.3847/1538-4357/ac1596}{{\em Astrophys. J.}
  {\bfseries 926} no.~1, (2022) 54},
  \href{http://arxiv.org/abs/2008.12619}{{\ttfamily arXiv:2008.12619
  [astro-ph.CO]}}.

\bibitem{CMB-S4:2022ght}
{\bfseries CMB-S4} Collaboration, K.~Abazajian {\em et~al.}, ``{Snowmass 2021
  CMB-S4 White Paper},'' \href{http://arxiv.org/abs/2203.08024}{{\ttfamily
  arXiv:2203.08024 [astro-ph.CO]}}.

\bibitem{Chang:2022lrw}
C.~L. Chang {\em et~al.}, ``{Report of the Topical Group on Cosmic Frontier 5
  Dark Energy and Cosmic Acceleration: Cosmic Dawn and Before for Snowmass
  2021},'' in {\em {Snowmass 2021}}.
\newblock 9, 2022.
\newblock \href{http://arxiv.org/abs/2209.08265}{{\ttfamily arXiv:2209.08265
  [hep-ex]}}.

\bibitem{Weinberg:2005vy}
S.~Weinberg, ``{Quantum contributions to cosmological correlations},''
  \href{http://dx.doi.org/10.1103/PhysRevD.72.043514}{{\em Phys. Rev. D}
  {\bfseries 72} (2005) 043514},
  \href{http://arxiv.org/abs/hep-th/0506236}{{\ttfamily arXiv:hep-th/0506236}}.

\bibitem{Weinberg:2006ac}
S.~Weinberg, ``{Quantum contributions to cosmological correlations. II. Can
  these corrections become large?},''
  \href{http://dx.doi.org/10.1103/PhysRevD.74.023508}{{\em Phys. Rev. D}
  {\bfseries 74} (2006) 023508},
  \href{http://arxiv.org/abs/hep-th/0605244}{{\ttfamily arXiv:hep-th/0605244}}.

\bibitem{Salopek:1990jq}
D.~S. Salopek and J.~R. Bond, ``{Nonlinear evolution of long wavelength metric
  fluctuations in inflationary models},''
  \href{http://dx.doi.org/10.1103/PhysRevD.42.3936}{{\em Phys. Rev. D}
  {\bfseries 42} (1990) 3936--3962}.

\bibitem{Maldacena:2002vr}
J.~M. Maldacena, ``{Non-Gaussian features of primordial fluctuations in single
  field inflationary models},''
  \href{http://dx.doi.org/10.1088/1126-6708/2003/05/013}{{\em JHEP} {\bfseries
  05} (2003) 013}, \href{http://arxiv.org/abs/astro-ph/0210603}{{\ttfamily
  arXiv:astro-ph/0210603}}.

\bibitem{Creminelli:2004yq}
P.~Creminelli and M.~Zaldarriaga, ``{Single field consistency relation for the
  3-point function},''
  \href{http://dx.doi.org/10.1088/1475-7516/2004/10/006}{{\em JCAP} {\bfseries
  10} (2004) 006}, \href{http://arxiv.org/abs/astro-ph/0407059}{{\ttfamily
  arXiv:astro-ph/0407059}}.

\bibitem{Alvarez:2014vva}
M.~Alvarez {\em et~al.}, ``{Testing Inflation with Large Scale Structure:
  Connecting Hopes with Reality},''
  \href{http://arxiv.org/abs/1412.4671}{{\ttfamily arXiv:1412.4671
  [astro-ph.CO]}}.

\bibitem{Weinberg:2003sw}
S.~Weinberg, ``{Adiabatic modes in cosmology},''
  \href{http://dx.doi.org/10.1103/PhysRevD.67.123504}{{\em Phys. Rev. D}
  {\bfseries 67} (2003) 123504},
  \href{http://arxiv.org/abs/astro-ph/0302326}{{\ttfamily
  arXiv:astro-ph/0302326}}.

\bibitem{Hinterbichler:2012nm}
K.~Hinterbichler, L.~Hui, and J.~Khoury, ``{Conformal Symmetries of Adiabatic
  Modes in Cosmology},''
  \href{http://dx.doi.org/10.1088/1475-7516/2012/08/017}{{\em JCAP} {\bfseries
  08} (2012) 017}, \href{http://arxiv.org/abs/1203.6351}{{\ttfamily
  arXiv:1203.6351 [hep-th]}}.

\bibitem{Hinterbichler:2013dpa}
K.~Hinterbichler, L.~Hui, and J.~Khoury, ``{An Infinite Set of Ward Identities
  for Adiabatic Modes in Cosmology},''
  \href{http://dx.doi.org/10.1088/1475-7516/2014/01/039}{{\em JCAP} {\bfseries
  01} (2014) 039}, \href{http://arxiv.org/abs/1304.5527}{{\ttfamily
  arXiv:1304.5527 [hep-th]}}.

\bibitem{Pimentel:2012tw}
G.~L. Pimentel, L.~Senatore, and M.~Zaldarriaga, ``{On Loops in Inflation III:
  Time Independence of zeta in Single Clock Inflation},''
  \href{http://dx.doi.org/10.1007/JHEP07(2012)166}{{\em JHEP} {\bfseries 07}
  (2012) 166}, \href{http://arxiv.org/abs/1203.6651}{{\ttfamily arXiv:1203.6651
  [hep-th]}}.

\bibitem{Assassi:2012et}
V.~Assassi, D.~Baumann, and D.~Green, ``{Symmetries and Loops in Inflation},''
  \href{http://dx.doi.org/10.1007/JHEP02(2013)151}{{\em JHEP} {\bfseries 02}
  (2013) 151}, \href{http://arxiv.org/abs/1210.7792}{{\ttfamily arXiv:1210.7792
  [hep-th]}}.

\bibitem{Senatore:2012ya}
L.~Senatore and M.~Zaldarriaga, ``{The constancy of $\zeta$ in single-clock
  Inflation at all loops},''
  \href{http://dx.doi.org/10.1007/JHEP09(2013)148}{{\em JHEP} {\bfseries 09}
  (2013) 148}, \href{http://arxiv.org/abs/1210.6048}{{\ttfamily arXiv:1210.6048
  [hep-th]}}.

\bibitem{Antoniadis:1985pj}
I.~Antoniadis, J.~Iliopoulos, and T.~N. Tomaras, ``{Quantum Instability of De
  Sitter Space},'' \href{http://dx.doi.org/10.1103/PhysRevLett.56.1319}{{\em
  Phys. Rev. Lett.} {\bfseries 56} (1986) 1319}.

\bibitem{Tsamis:1994ca}
N.~C. Tsamis and R.~P. Woodard, ``{Strong infrared effects in quantum
  gravity},'' \href{http://dx.doi.org/10.1006/aphy.1995.1015}{{\em Annals
  Phys.} {\bfseries 238} (1995) 1--82}.

\bibitem{Tsamis:1996qm}
N.~C. Tsamis and R.~P. Woodard, ``{The Quantum gravitational back reaction on
  inflation},'' \href{http://dx.doi.org/10.1006/aphy.1997.5613}{{\em Annals
  Phys.} {\bfseries 253} (1997) 1--54},
  \href{http://arxiv.org/abs/hep-ph/9602316}{{\ttfamily arXiv:hep-ph/9602316}}.

\bibitem{Adshead:2008gk}
P.~Adshead, R.~Easther, and E.~A. Lim, ``{Cosmology With Many Light Scalar
  Fields: Stochastic Inflation and Loop Corrections},''
  \href{http://dx.doi.org/10.1103/PhysRevD.79.063504}{{\em Phys. Rev. D}
  {\bfseries 79} (2009) 063504},
  \href{http://arxiv.org/abs/0809.4008}{{\ttfamily arXiv:0809.4008 [hep-th]}}.

\bibitem{Senatore:2009cf}
L.~Senatore and M.~Zaldarriaga, ``{On Loops in Inflation},''
  \href{http://dx.doi.org/10.1007/JHEP12(2010)008}{{\em JHEP} {\bfseries 12}
  (2010) 008}, \href{http://arxiv.org/abs/0912.2734}{{\ttfamily arXiv:0912.2734
  [hep-th]}}.

\bibitem{Burgess:2010dd}
C.~P. Burgess, R.~Holman, L.~Leblond, and S.~Shandera, ``{Breakdown of
  Semiclassical Methods in de Sitter Space},''
  \href{http://dx.doi.org/10.1088/1475-7516/2010/10/017}{{\em JCAP} {\bfseries
  10} (2010) 017}, \href{http://arxiv.org/abs/1005.3551}{{\ttfamily
  arXiv:1005.3551 [hep-th]}}.

\bibitem{Senatore:2012nq}
L.~Senatore and M.~Zaldarriaga, ``{On Loops in Inflation II: IR Effects in
  Single Clock Inflation},''
  \href{http://dx.doi.org/10.1007/JHEP01(2013)109}{{\em JHEP} {\bfseries 01}
  (2013) 109}, \href{http://arxiv.org/abs/1203.6354}{{\ttfamily arXiv:1203.6354
  [hep-th]}}.

\bibitem{Akhmedov:2013vka}
E.~T. Akhmedov, ``{Lecture notes on interacting quantum fields in de Sitter
  space},'' \href{http://dx.doi.org/10.1142/S0218271814300018}{{\em Int. J.
  Mod. Phys. D} {\bfseries 23} (2014) 1430001},
  \href{http://arxiv.org/abs/1309.2557}{{\ttfamily arXiv:1309.2557 [hep-th]}}.

\bibitem{Anninos:2014lwa}
D.~Anninos, T.~Anous, D.~Z. Freedman, and G.~Konstantinidis, ``{Late-time
  Structure of the Bunch-Davies De Sitter Wavefunction},''
  \href{http://dx.doi.org/10.1088/1475-7516/2015/11/048}{{\em JCAP} {\bfseries
  11} (2015) 048}, \href{http://arxiv.org/abs/1406.5490}{{\ttfamily
  arXiv:1406.5490 [hep-th]}}.

\bibitem{Akhmedov:2019cfd}
E.~T. Akhmedov, U.~Moschella, and F.~K. Popov, ``{Characters of different
  secular effects in various patches of de Sitter space},''
  \href{http://dx.doi.org/10.1103/PhysRevD.99.086009}{{\em Phys. Rev. D}
  {\bfseries 99} no.~8, (2019) 086009},
  \href{http://arxiv.org/abs/1901.07293}{{\ttfamily arXiv:1901.07293
  [hep-th]}}.

\bibitem{Gorbenko:2019rza}
V.~Gorbenko and L.~Senatore, ``{$\lambda \phi^4$ in dS},''
  \href{http://arxiv.org/abs/1911.00022}{{\ttfamily arXiv:1911.00022
  [hep-th]}}.

\bibitem{Mirbabayi:2019qtx}
M.~Mirbabayi, ``{Infrared dynamics of a light scalar field in de Sitter},''
  \href{http://dx.doi.org/10.1088/1475-7516/2020/12/006}{{\em JCAP} {\bfseries
  12} (2020) 006}, \href{http://arxiv.org/abs/1911.00564}{{\ttfamily
  arXiv:1911.00564 [hep-th]}}.

\bibitem{Baumgart:2019clc}
M.~Baumgart and R.~Sundrum, ``{De Sitter Diagrammar and the Resummation of
  Time},'' \href{http://dx.doi.org/10.1007/JHEP07(2020)119}{{\em JHEP}
  {\bfseries 07} (2020) 119}, \href{http://arxiv.org/abs/1912.09502}{{\ttfamily
  arXiv:1912.09502 [hep-th]}}.

\bibitem{Baumgart:2020oby}
M.~Baumgart and R.~Sundrum, ``{Manifestly Causal In-In Perturbation Theory
  about the Interacting Vacuum},''
  \href{http://dx.doi.org/10.1007/JHEP03(2021)080}{{\em JHEP} {\bfseries 03}
  (2021) 080}, \href{http://arxiv.org/abs/2010.10785}{{\ttfamily
  arXiv:2010.10785 [hep-th]}}.

\bibitem{Benincasa:2022gtd}
P.~Benincasa, ``{Amplitudes meet Cosmology: A (Scalar) Primer},''
  \href{http://arxiv.org/abs/2203.15330}{{\ttfamily arXiv:2203.15330
  [hep-th]}}.

\bibitem{Benincasa:2024ptf}
P.~Benincasa, G.~Brunello, M.~K. Mandal, P.~Mastrolia, and F.~Vaz\~ao, ``{On
  one-loop corrections to the Bunch-Davies wavefunction of the universe},''
  \href{http://arxiv.org/abs/2408.16386}{{\ttfamily arXiv:2408.16386
  [hep-th]}}.

\bibitem{Manohar:2018aog}
A.~V. Manohar, ``{Introduction to Effective Field Theories},''
  \href{http://arxiv.org/abs/1804.05863}{{\ttfamily arXiv:1804.05863
  [hep-ph]}}.

\bibitem{Cohen:2019wxr}
T.~Cohen, ``{As Scales Become Separated: Lectures on Effective Field Theory},''
  {\em PoS} {\bfseries TASI2018} (2019) 011,
  \href{http://arxiv.org/abs/1903.03622}{{\ttfamily arXiv:1903.03622
  [hep-ph]}}.

\bibitem{Green2023}
D.~Green, {\em EFT for de Sitter Space}, pp.~1--32.
\newblock Springer Nature Singapore, Singapore, 2023.
\newblock \href{http://arxiv.org/abs/2210.05820}{{\ttfamily arXiv:2210.05820
  [hep-th]}}.

\bibitem{Cohen:2020php}
T.~Cohen and D.~Green, ``{Soft de Sitter Effective Theory},''
  \href{http://dx.doi.org/10.1007/JHEP12(2020)041}{{\em JHEP} {\bfseries 12}
  (2020) 041}, \href{http://arxiv.org/abs/2007.03693}{{\ttfamily
  arXiv:2007.03693 [hep-th]}}.

\bibitem{Starobinsky:1986fx}
A.~A. Starobinsky, ``{STOCHASTIC DE SITTER (INFLATIONARY) STAGE IN THE EARLY
  UNIVERSE},'' \href{http://dx.doi.org/10.1007/3-540-16452-9_6}{{\em Lect.
  Notes Phys.} {\bfseries 246} (1986) 107--126}.

\bibitem{Nambu:1987ef}
Y.~Nambu and M.~Sasaki, ``{Stochastic Stage of an Inflationary Universe
  Model},'' \href{http://dx.doi.org/10.1016/0370-2693(88)90974-4}{{\em Phys.
  Lett. B} {\bfseries 205} (1988) 441--446}.

\bibitem{Starobinsky:1994bd}
A.~A. Starobinsky and J.~Yokoyama, ``{Equilibrium state of a selfinteracting
  scalar field in the De Sitter background},''
  \href{http://dx.doi.org/10.1103/PhysRevD.50.6357}{{\em Phys. Rev. D}
  {\bfseries 50} (1994) 6357--6368},
  \href{http://arxiv.org/abs/astro-ph/9407016}{{\ttfamily
  arXiv:astro-ph/9407016}}.

\bibitem{Cohen:2021fzf}
T.~Cohen, D.~Green, A.~Premkumar, and A.~Ridgway, ``{Stochastic Inflation at
  NNLO},'' \href{http://dx.doi.org/10.1007/JHEP09(2021)159}{{\em JHEP}
  {\bfseries 09} (2021) 159}, \href{http://arxiv.org/abs/2106.09728}{{\ttfamily
  arXiv:2106.09728 [hep-th]}}.

\bibitem{Cohen:2021jbo}
T.~Cohen, D.~Green, and A.~Premkumar, ``{A tail of eternal inflation},''
  \href{http://dx.doi.org/10.21468/SciPostPhys.14.5.109}{{\em SciPost Phys.}
  {\bfseries 14} no.~5, (2023) 109},
  \href{http://arxiv.org/abs/2111.09332}{{\ttfamily arXiv:2111.09332
  [hep-th]}}.

\bibitem{Cohen:2022clv}
T.~Cohen, D.~Green, and A.~Premkumar, ``{Large deviations in the early
  Universe},'' \href{http://dx.doi.org/10.1103/PhysRevD.107.083501}{{\em Phys.
  Rev. D} {\bfseries 107} no.~8, (2023) 083501},
  \href{http://arxiv.org/abs/2212.02535}{{\ttfamily arXiv:2212.02535
  [hep-th]}}.

\bibitem{Cheung:2007st}
C.~Cheung, P.~Creminelli, A.~L. Fitzpatrick, J.~Kaplan, and L.~Senatore, ``{The
  Effective Field Theory of Inflation},''
  \href{http://dx.doi.org/10.1088/1126-6708/2008/03/014}{{\em JHEP} {\bfseries
  03} (2008) 014}, \href{http://arxiv.org/abs/0709.0293}{{\ttfamily
  arXiv:0709.0293 [hep-th]}}.

\bibitem{Green:2020ebl}
D.~Green and E.~Pajer, ``{On the Symmetries of Cosmological Perturbations},''
  \href{http://dx.doi.org/10.1088/1475-7516/2020/09/032}{{\em JCAP} {\bfseries
  09} (2020) 032}, \href{http://arxiv.org/abs/2004.09587}{{\ttfamily
  arXiv:2004.09587 [hep-th]}}.

\bibitem{Creminelli:2012ed}
P.~Creminelli, J.~Nore\~na, and M.~Simonovi\'c, ``{Conformal consistency
  relations for single-field inflation},''
  \href{http://dx.doi.org/10.1088/1475-7516/2012/07/052}{{\em JCAP} {\bfseries
  07} (2012) 052}, \href{http://arxiv.org/abs/1203.4595}{{\ttfamily
  arXiv:1203.4595 [hep-th]}}.

\bibitem{Assassi:2012zq}
V.~Assassi, D.~Baumann, and D.~Green, ``{On Soft Limits of Inflationary
  Correlation Functions},''
  \href{http://dx.doi.org/10.1088/1475-7516/2012/11/047}{{\em JCAP} {\bfseries
  11} (2012) 047}, \href{http://arxiv.org/abs/1204.4207}{{\ttfamily
  arXiv:1204.4207 [hep-th]}}.

\bibitem{Goldberger:2013rsa}
W.~D. Goldberger, L.~Hui, and A.~Nicolis, ``{One-particle-irreducible
  consistency relations for cosmological perturbations},''
  \href{http://dx.doi.org/10.1103/PhysRevD.87.103520}{{\em Phys. Rev. D}
  {\bfseries 87} no.~10, (2013) 103520},
  \href{http://arxiv.org/abs/1303.1193}{{\ttfamily arXiv:1303.1193 [hep-th]}}.

\bibitem{Pimentel:2013gza}
G.~L. Pimentel, ``{Inflationary Consistency Conditions from a Wavefunctional
  Perspective},'' \href{http://dx.doi.org/10.1007/JHEP02(2014)124}{{\em JHEP}
  {\bfseries 02} (2014) 124}, \href{http://arxiv.org/abs/1309.1793}{{\ttfamily
  arXiv:1309.1793 [hep-th]}}.

\bibitem{Chen:2009zp}
X.~Chen and Y.~Wang, ``{Quasi-Single Field Inflation and Non-Gaussianities},''
  \href{http://dx.doi.org/10.1088/1475-7516/2010/04/027}{{\em JCAP} {\bfseries
  04} (2010) 027}, \href{http://arxiv.org/abs/0911.3380}{{\ttfamily
  arXiv:0911.3380 [hep-th]}}.

\bibitem{Baumann:2011nk}
D.~Baumann and D.~Green, ``{Signatures of Supersymmetry from the Early
  Universe},'' \href{http://dx.doi.org/10.1103/PhysRevD.85.103520}{{\em Phys.
  Rev. D} {\bfseries 85} (2012) 103520},
  \href{http://arxiv.org/abs/1109.0292}{{\ttfamily arXiv:1109.0292 [hep-th]}}.

\bibitem{Noumi:2012vr}
T.~Noumi, M.~Yamaguchi, and D.~Yokoyama, ``{Effective field theory approach to
  quasi-single field inflation and effects of heavy fields},''
  \href{http://dx.doi.org/10.1007/JHEP06(2013)051}{{\em JHEP} {\bfseries 06}
  (2013) 051}, \href{http://arxiv.org/abs/1211.1624}{{\ttfamily arXiv:1211.1624
  [hep-th]}}.

\bibitem{Polchinski:1983gv}
J.~Polchinski, ``{Renormalization and Effective Lagrangians},''
  \href{http://dx.doi.org/10.1016/0550-3213(84)90287-6}{{\em Nucl. Phys. B}
  {\bfseries 231} (1984) 269--295}.

\bibitem{Sleight:2019mgd}
C.~Sleight, ``{A Mellin Space Approach to Cosmological Correlators},''
  \href{http://dx.doi.org/10.1007/JHEP01(2020)090}{{\em JHEP} {\bfseries 01}
  (2020) 090}, \href{http://arxiv.org/abs/1906.12302}{{\ttfamily
  arXiv:1906.12302 [hep-th]}}.

\bibitem{Sleight:2019hfp}
C.~Sleight and M.~Taronna, ``{Bootstrapping Inflationary Correlators in Mellin
  Space},'' \href{http://dx.doi.org/10.1007/JHEP02(2020)098}{{\em JHEP}
  {\bfseries 02} (2020) 098}, \href{http://arxiv.org/abs/1907.01143}{{\ttfamily
  arXiv:1907.01143 [hep-th]}}.

\bibitem{Premkumar:2021mlz}
A.~Premkumar, ``{Regulating loops in de Sitter spacetime},''
  \href{http://dx.doi.org/10.1103/PhysRevD.109.045003}{{\em Phys. Rev. D}
  {\bfseries 109} no.~4, (2024) 045003},
  \href{http://arxiv.org/abs/2110.12504}{{\ttfamily arXiv:2110.12504
  [hep-th]}}.

\bibitem{Qin:2022lva}
Z.~Qin and Z.-Z. Xianyu, ``{Phase information in cosmological collider
  signals},'' \href{http://dx.doi.org/10.1007/JHEP10(2022)192}{{\em JHEP}
  {\bfseries 10} (2022) 192}, \href{http://arxiv.org/abs/2205.01692}{{\ttfamily
  arXiv:2205.01692 [hep-th]}}.

\bibitem{Qin:2023bjk}
Z.~Qin and Z.-Z. Xianyu, ``{Inflation correlators at the one-loop order:
  nonanalyticity, factorization, cutting rule, and OPE},''
  \href{http://dx.doi.org/10.1007/JHEP09(2023)116}{{\em JHEP} {\bfseries 09}
  (2023) 116}, \href{http://arxiv.org/abs/2304.13295}{{\ttfamily
  arXiv:2304.13295 [hep-th]}}.

\bibitem{Cohen:2024anu}
T.~Cohen, D.~Green, and Y.~Huang, ``{Operator Origin of Anomalous Dimensions in
  de Sitter Space},'' \href{http://arxiv.org/abs/2407.08581}{{\ttfamily
  arXiv:2407.08581 [hep-th]}}.

\bibitem{Bros:2010rku}
J.~Bros, H.~Epstein, and U.~Moschella, ``{Particle decays and stability on the
  de Sitter universe},''
  \href{http://dx.doi.org/10.1007/s00023-010-0042-7}{{\em Annales Henri
  Poincare} {\bfseries 11} (2010) 611--658},
  \href{http://arxiv.org/abs/0812.3513}{{\ttfamily arXiv:0812.3513 [hep-th]}}.

\bibitem{Marolf:2010zp}
D.~Marolf and I.~A. Morrison, ``{The IR stability of de Sitter: Loop
  corrections to scalar propagators},''
  \href{http://dx.doi.org/10.1103/PhysRevD.82.105032}{{\em Phys. Rev. D}
  {\bfseries 82} (2010) 105032},
  \href{http://arxiv.org/abs/1006.0035}{{\ttfamily arXiv:1006.0035 [gr-qc]}}.

\bibitem{Hogervorst:2021uvp}
M.~Hogervorst, J.~a. Penedones, and K.~S. Vaziri, ``{Towards the
  non-perturbative cosmological bootstrap},''
  \href{http://dx.doi.org/10.1007/JHEP02(2023)162}{{\em JHEP} {\bfseries 02}
  (2023) 162}, \href{http://arxiv.org/abs/2107.13871}{{\ttfamily
  arXiv:2107.13871 [hep-th]}}.

\bibitem{Chakraborty:2023qbp}
P.~Chakraborty and J.~Stout, ``{Light scalars at the cosmological collider},''
  \href{http://dx.doi.org/10.1007/JHEP02(2024)021}{{\em JHEP} {\bfseries 02}
  (2024) 021}, \href{http://arxiv.org/abs/2310.01494}{{\ttfamily
  arXiv:2310.01494 [hep-th]}}.

\bibitem{Chakraborty:2023eoq}
P.~Chakraborty and J.~Stout, ``{Compact scalars at the cosmological
  collider},'' \href{http://dx.doi.org/10.1007/JHEP03(2024)149}{{\em JHEP}
  {\bfseries 03} (2024) 149}, \href{http://arxiv.org/abs/2311.09219}{{\ttfamily
  arXiv:2311.09219 [hep-th]}}.

\bibitem{Green:2023ids}
D.~Green, Y.~Huang, C.-H. Shen, and D.~Baumann, ``{Positivity from Cosmological
  Correlators},'' \href{http://dx.doi.org/10.1007/JHEP04(2024)034}{{\em JHEP}
  {\bfseries 04} (2024) 034}, \href{http://arxiv.org/abs/2310.02490}{{\ttfamily
  arXiv:2310.02490 [hep-th]}}.

\bibitem{Creminelli:2006xe}
P.~Creminelli, M.~A. Luty, A.~Nicolis, and L.~Senatore, ``{Starting the
  Universe: Stable Violation of the Null Energy Condition and Non-standard
  Cosmologies},'' \href{http://dx.doi.org/10.1088/1126-6708/2006/12/080}{{\em
  JHEP} {\bfseries 12} (2006) 080},
  \href{http://arxiv.org/abs/hep-th/0606090}{{\ttfamily arXiv:hep-th/0606090}}.

\bibitem{Baumann:2011su}
D.~Baumann and D.~Green, ``{Equilateral Non-Gaussianity and New Physics on the
  Horizon},'' \href{http://dx.doi.org/10.1088/1475-7516/2011/09/014}{{\em JCAP}
  {\bfseries 09} (2011) 014}, \href{http://arxiv.org/abs/1102.5343}{{\ttfamily
  arXiv:1102.5343 [hep-th]}}.

\bibitem{Green:2022slj}
D.~Green, Y.~Huang, and C.-H. Shen, ``{Inflationary Adler conditions},''
  \href{http://dx.doi.org/10.1103/PhysRevD.107.043534}{{\em Phys. Rev. D}
  {\bfseries 107} no.~4, (2023) 043534},
  \href{http://arxiv.org/abs/2208.14544}{{\ttfamily arXiv:2208.14544
  [hep-th]}}.

\bibitem{Green:2024hbw}
D.~Green, K.~Gupta, and Y.~Huang, ``{A Goldstone boson equivalence for
  inflation},'' \href{http://dx.doi.org/10.1007/JHEP09(2024)117}{{\em JHEP}
  {\bfseries 09} (2024) 117}, \href{http://arxiv.org/abs/2403.05274}{{\ttfamily
  arXiv:2403.05274 [hep-th]}}.

\bibitem{Salopek:1990re}
D.~S. Salopek and J.~R. Bond, ``{Stochastic inflation and nonlinear gravity},''
  \href{http://dx.doi.org/10.1103/PhysRevD.43.1005}{{\em Phys. Rev. D}
  {\bfseries 43} (1991) 1005--1031}.

\bibitem{Wands:2000dp}
D.~Wands, K.~A. Malik, D.~H. Lyth, and A.~R. Liddle, ``{A New approach to the
  evolution of cosmological perturbations on large scales},''
  \href{http://dx.doi.org/10.1103/PhysRevD.62.043527}{{\em Phys. Rev. D}
  {\bfseries 62} (2000) 043527},
  \href{http://arxiv.org/abs/astro-ph/0003278}{{\ttfamily
  arXiv:astro-ph/0003278}}.

\bibitem{Chen:2006nt}
X.~Chen, M.-x. Huang, S.~Kachru, and G.~Shiu, ``{Observational signatures and
  non-Gaussianities of general single field inflation},''
  \href{http://dx.doi.org/10.1088/1475-7516/2007/01/002}{{\em JCAP} {\bfseries
  01} (2007) 002}, \href{http://arxiv.org/abs/hep-th/0605045}{{\ttfamily
  arXiv:hep-th/0605045}}.

\bibitem{Cheung:2007sv}
C.~Cheung, A.~L. Fitzpatrick, J.~Kaplan, and L.~Senatore, ``{On the consistency
  relation of the 3-point function in single field inflation},''
  \href{http://dx.doi.org/10.1088/1475-7516/2008/02/021}{{\em JCAP} {\bfseries
  02} (2008) 021}, \href{http://arxiv.org/abs/0709.0295}{{\ttfamily
  arXiv:0709.0295 [hep-th]}}.

\bibitem{Baumann:2014cja}
D.~Baumann, D.~Green, and R.~A. Porto, ``{B-modes and the Nature of
  Inflation},'' \href{http://dx.doi.org/10.1088/1475-7516/2015/01/016}{{\em
  JCAP} {\bfseries 01} (2015) 016},
  \href{http://arxiv.org/abs/1407.2621}{{\ttfamily arXiv:1407.2621 [hep-th]}}.

\bibitem{Green:toappear}
D.~Green, K.~Gupta, and G.~Sun, ``{in preparation},''.

\bibitem{Creminelli:2008es}
P.~Creminelli, S.~Dubovsky, A.~Nicolis, L.~Senatore, and M.~Zaldarriaga, ``{The
  Phase Transition to Slow-roll Eternal Inflation},''
  \href{http://dx.doi.org/10.1088/1126-6708/2008/09/036}{{\em JHEP} {\bfseries
  09} (2008) 036}, \href{http://arxiv.org/abs/0802.1067}{{\ttfamily
  arXiv:0802.1067 [hep-th]}}.

\bibitem{Dubovsky:2008rf}
S.~Dubovsky, L.~Senatore, and G.~Villadoro, ``{The Volume of the Universe after
  Inflation and de Sitter Entropy},''
  \href{http://dx.doi.org/10.1088/1126-6708/2009/04/118}{{\em JHEP} {\bfseries
  04} (2009) 118}, \href{http://arxiv.org/abs/0812.2246}{{\ttfamily
  arXiv:0812.2246 [hep-th]}}.

\bibitem{Dubovsky:2011uy}
S.~Dubovsky, L.~Senatore, and G.~Villadoro, ``{Universality of the Volume Bound
  in Slow-Roll Eternal Inflation},''
  \href{http://dx.doi.org/10.1007/JHEP05(2012)035}{{\em JHEP} {\bfseries 05}
  (2012) 035}, \href{http://arxiv.org/abs/1111.1725}{{\ttfamily arXiv:1111.1725
  [hep-th]}}.

\bibitem{Lewandowski:2013aka}
M.~Lewandowski and A.~Perko, ``{Leading slow roll corrections to the volume of
  the universe and the entropy bound},''
  \href{http://dx.doi.org/10.1007/JHEP12(2014)060}{{\em JHEP} {\bfseries 12}
  (2014) 060}, \href{http://arxiv.org/abs/1309.6705}{{\ttfamily arXiv:1309.6705
  [hep-th]}}.

\bibitem{Chaicherdsakul:2006ui}
K.~Chaicherdsakul, ``{Quantum Cosmological Correlations in an Inflating
  Universe: Can fermion and gauge fields loops give a scale free spectrum?},''
  \href{http://dx.doi.org/10.1103/PhysRevD.75.063522}{{\em Phys. Rev. D}
  {\bfseries 75} (2007) 063522},
  \href{http://arxiv.org/abs/hep-th/0611352}{{\ttfamily arXiv:hep-th/0611352}}.

\bibitem{Seery:2007we}
D.~Seery, ``{One-loop corrections to a scalar field during inflation},''
  \href{http://dx.doi.org/10.1088/1475-7516/2007/11/025}{{\em JCAP} {\bfseries
  11} (2007) 025}, \href{http://arxiv.org/abs/0707.3377}{{\ttfamily
  arXiv:0707.3377 [astro-ph]}}.

\bibitem{Dimastrogiovanni:2008af}
E.~Dimastrogiovanni and N.~Bartolo, ``{One-loop graviton corrections to the
  curvature perturbation from inflation},''
  \href{http://dx.doi.org/10.1088/1475-7516/2008/11/016}{{\em JCAP} {\bfseries
  11} (2008) 016}, \href{http://arxiv.org/abs/0807.2790}{{\ttfamily
  arXiv:0807.2790 [astro-ph]}}.

\bibitem{Kristiano:2022zpn}
J.~Kristiano and J.~Yokoyama, ``{Perturbative region on non-Gaussian parameter
  space in single-field inflation},''
  \href{http://dx.doi.org/10.1088/1475-7516/2022/07/007}{{\em JCAP} {\bfseries
  07} no.~07, (2022) 007}, \href{http://arxiv.org/abs/2204.05202}{{\ttfamily
  arXiv:2204.05202 [hep-th]}}.

\bibitem{Seljak:1996gy}
U.~Seljak and M.~Zaldarriaga, ``{Signature of gravity waves in polarization of
  the microwave background},''
  \href{http://dx.doi.org/10.1103/PhysRevLett.78.2054}{{\em Phys. Rev. Lett.}
  {\bfseries 78} (1997) 2054--2057},
  \href{http://arxiv.org/abs/astro-ph/9609169}{{\ttfamily
  arXiv:astro-ph/9609169}}.

\bibitem{Kamionkowski:1996ks}
M.~Kamionkowski, A.~Kosowsky, and A.~Stebbins, ``{Statistics of cosmic
  microwave background polarization},''
  \href{http://dx.doi.org/10.1103/PhysRevD.55.7368}{{\em Phys. Rev. D}
  {\bfseries 55} (1997) 7368--7388},
  \href{http://arxiv.org/abs/astro-ph/9611125}{{\ttfamily
  arXiv:astro-ph/9611125}}.

\bibitem{Maldacena:2011nz}
J.~M. Maldacena and G.~L. Pimentel, ``{On graviton non-Gaussianities during
  inflation},'' \href{http://dx.doi.org/10.1007/JHEP09(2011)045}{{\em JHEP}
  {\bfseries 09} (2011) 045}, \href{http://arxiv.org/abs/1104.2846}{{\ttfamily
  arXiv:1104.2846 [hep-th]}}.

\bibitem{Meerburg:2016ecv}
P.~D. Meerburg, J.~Meyers, A.~van Engelen, and Y.~Ali-Ha\"\i{}moud, ``{CMB B
  -mode non-Gaussianity},''
  \href{http://dx.doi.org/10.1103/PhysRevD.93.123511}{{\em Phys. Rev. D}
  {\bfseries 93} (2016) 123511},
  \href{http://arxiv.org/abs/1603.02243}{{\ttfamily arXiv:1603.02243
  [astro-ph.CO]}}.

\bibitem{Dimastrogiovanni:2018uqy}
E.~Dimastrogiovanni, M.~Fasiello, and G.~Tasinato, ``{Probing the inflationary
  particle content: extra spin-2 field},''
  \href{http://dx.doi.org/10.1088/1475-7516/2018/08/016}{{\em JCAP} {\bfseries
  08} (2018) 016}, \href{http://arxiv.org/abs/1806.00850}{{\ttfamily
  arXiv:1806.00850 [astro-ph.CO]}}.

\bibitem{Bordin:2020eui}
L.~Bordin and G.~Cabass, ``{Graviton non-Gaussianities and Parity Violation in
  the EFT of Inflation},''
  \href{http://dx.doi.org/10.1088/1475-7516/2020/07/014}{{\em JCAP} {\bfseries
  07} (2020) 014}, \href{http://arxiv.org/abs/2004.00619}{{\ttfamily
  arXiv:2004.00619 [astro-ph.CO]}}.

\bibitem{Cabass:2021iii}
G.~Cabass, ``{Zoology of graviton non-Gaussianities},''
  \href{http://dx.doi.org/10.1088/1475-7516/2021/12/001}{{\em JCAP} {\bfseries
  12} no.~12, (2021) 001}, \href{http://arxiv.org/abs/2103.09816}{{\ttfamily
  arXiv:2103.09816 [hep-th]}}.

\bibitem{Cabass:2021fnw}
G.~Cabass, E.~Pajer, D.~Stefanyszyn, and J.~Supel, ``{Bootstrapping large
  graviton non-Gaussianities},''
  \href{http://dx.doi.org/10.1007/JHEP05(2022)077}{{\em JHEP} {\bfseries 05}
  (2022) 077}, \href{http://arxiv.org/abs/2109.10189}{{\ttfamily
  arXiv:2109.10189 [hep-th]}}.

\bibitem{Cabass:2022jda}
G.~Cabass, D.~Stefanyszyn, J.~Supel, and A.~Thavanesan, ``{On graviton
  non-Gaussianities in the Effective Field Theory of Inflation},''
  \href{http://dx.doi.org/10.1007/JHEP10(2022)154}{{\em JHEP} {\bfseries 10}
  (2022) 154}, \href{http://arxiv.org/abs/2209.00677}{{\ttfamily
  arXiv:2209.00677 [hep-th]}}.

\bibitem{Creminelli:2014wna}
P.~Creminelli, J.~Gleyzes, J.~Nore\~na, and F.~Vernizzi, ``{Resilience of the
  standard predictions for primordial tensor modes},''
  \href{http://dx.doi.org/10.1103/PhysRevLett.113.231301}{{\em Phys. Rev.
  Lett.} {\bfseries 113} no.~23, (2014) 231301},
  \href{http://arxiv.org/abs/1407.8439}{{\ttfamily arXiv:1407.8439
  [astro-ph.CO]}}.

\bibitem{Bordin:2016ruc}
L.~Bordin, P.~Creminelli, M.~Mirbabayi, and J.~Nore\~na, ``{Tensor Squeezed
  Limits and the Higuchi Bound},''
  \href{http://dx.doi.org/10.1088/1475-7516/2016/09/041}{{\em JCAP} {\bfseries
  09} (2016) 041}, \href{http://arxiv.org/abs/1605.08424}{{\ttfamily
  arXiv:1605.08424 [astro-ph.CO]}}.

\bibitem{Flauger:2022hie}
R.~Flauger, V.~Gorbenko, A.~Joyce, L.~McAllister, G.~Shiu, and E.~Silverstein,
  ``{Snowmass White Paper: Cosmology at the Theory Frontier},'' in {\em
  {Snowmass 2021}}.
\newblock 3, 2022.
\newblock \href{http://arxiv.org/abs/2203.07629}{{\ttfamily arXiv:2203.07629
  [hep-th]}}.

\bibitem{Vilenkin:1983xq}
A.~Vilenkin, ``{The Birth of Inflationary Universes},''
  \href{http://dx.doi.org/10.1103/PhysRevD.27.2848}{{\em Phys. Rev. D}
  {\bfseries 27} (1983) 2848}.

\bibitem{Linde:1986fd}
A.~D. Linde, ``{Eternally Existing Selfreproducing Chaotic Inflationary
  Universe},'' \href{http://dx.doi.org/10.1016/0370-2693(86)90611-8}{{\em Phys.
  Lett. B} {\bfseries 175} (1986) 395--400}.

\bibitem{Arkani-Hamed:2007ryv}
N.~Arkani-Hamed, S.~Dubovsky, A.~Nicolis, E.~Trincherini, and G.~Villadoro,
  ``{A Measure of de Sitter entropy and eternal inflation},''
  \href{http://dx.doi.org/10.1088/1126-6708/2007/05/055}{{\em JHEP} {\bfseries
  05} (2007) 055}, \href{http://arxiv.org/abs/0704.1814}{{\ttfamily
  arXiv:0704.1814 [hep-th]}}.

\bibitem{Hartle:1983ai}
J.~B. Hartle and S.~W. Hawking, ``{Wave Function of the Universe},''
  \href{http://dx.doi.org/10.1103/PhysRevD.28.2960}{{\em Phys. Rev. D}
  {\bfseries 28} (1983) 2960--2975}.

\bibitem{Maldacena:2024uhs}
J.~Maldacena, ``{Comments on the no boundary wavefunction and slow roll
  inflation},'' \href{http://arxiv.org/abs/2403.10510}{{\ttfamily
  arXiv:2403.10510 [hep-th]}}.

\bibitem{Panagopoulos:2020sxp}
G.~Panagopoulos and E.~Silverstein, ``{Multipoint correlators in multifield
  cosmology},'' \href{http://arxiv.org/abs/2003.05883}{{\ttfamily
  arXiv:2003.05883 [hep-th]}}.

\bibitem{Celoria:2021vjw}
M.~Celoria, P.~Creminelli, G.~Tambalo, and V.~Yingcharoenrat, ``{Beyond
  perturbation theory in inflation},''
  \href{http://dx.doi.org/10.1088/1475-7516/2021/06/051}{{\em JCAP} {\bfseries
  06} (2021) 051}, \href{http://arxiv.org/abs/2103.09244}{{\ttfamily
  arXiv:2103.09244 [hep-th]}}.

\bibitem{Creminelli:2024cge}
P.~Creminelli, S.~Renaux-Petel, G.~Tambalo, and V.~Yingcharoenrat,
  ``{Non-perturbative wavefunction of the universe in inflation with (resonant)
  features},'' \href{http://dx.doi.org/10.1007/JHEP03(2024)010}{{\em JHEP}
  {\bfseries 03} (2024) 010}, \href{http://arxiv.org/abs/2401.10212}{{\ttfamily
  arXiv:2401.10212 [hep-th]}}.

\bibitem{Panagopoulos:2019ail}
G.~Panagopoulos and E.~Silverstein, ``{Primordial Black Holes from non-Gaussian
  tails},'' \href{http://arxiv.org/abs/1906.02827}{{\ttfamily arXiv:1906.02827
  [hep-th]}}.

\end{thebibliography}\endgroup

\end{document}